\documentclass{aa}  

\usepackage{graphicx}
\usepackage{txfonts}
\usepackage{textcomp}

\usepackage{verbatim} 
\usepackage[normalem]{ulem}

\usepackage{xcolor}
\usepackage{soul}
\usepackage{bm}
\usepackage{lscape}

\def\smallskipt{\noalign {\smallskip}}
\def\co2{CO$_2$}
\def\ch4{CH$_4$}
\def\n2{H$_{2}$}
\def\h2{H$_{2}$}
\def\h2o{H$_2$O}
\def\nh3{NH$_3$}

\def\cm1{cm$^{-1}$}

\def\kvt{$k_{vt}$}
\def\tvib{$T_{vib}$}
\def\kg{$k_{g}$}
\def\up#1{$^{#1}$} 
\def\v#1{$v_{#1}$} 
\def\runits{W/(m$^2$\,sr\,$\mu$m)}

\def\flecha{$\rightarrow$}
\def\df{$\leftrightharpoons$}
\def\um{$\mu$m}
\def\cms{cm$^3$s$^{-1}$}
\def\deg{$^{\circ}$}
\def\kmin{k$_{min}$}
\def\kmax{k$_{max}$}

\begin{document} 

\title{The \ch4 abundance in Jupiter's upper atmosphere}
%
%
\titlerunning{The \ch4 abundance in Jupiter's upper atmosphere}
\author{A.\,S\'anchez-L\'opez\inst{1}, M.\,L\'opez-Puertas\inst{2}, M.\,Garc\'ia-Comas\inst{2}, B.\,Funke\inst{2}, T.\,Fouchet\inst{3}, \\
and I.\,A.\,G.\,Snellen\inst{1}}
\institute{Leiden Observatory, Leiden University, Postbus 9513, 2300 RA, Leiden, The Netherlands\\
\email{alexsl@strw.leidenuniv.nl}
\and
Instituto de Astrof{\'i}sica de Andaluc{\'i}a (IAA-CSIC), Glorieta de la Astronom{\'i}a s/n, 18008 Granada, Spain
\and
LESIA, Observatoire de Paris, Universit\'e PSL, CNRS, Sorbonne Universit\'e, Universit\'e de Paris, 5 place Jules Janssen, 92195 Meudon, France
}
\authorrunning{A. S\'anchez-L\'opez et al.}

\date{}

 
\abstract
{Hydrocarbon species, and in particular \ch4, play a key role in the stratosphere--thermosphere boundary of Jupiter, which occurs around the $\mu$-bar pressure level.
Previous analyses of solar occultation,  He and Ly-$\alpha$ airglow, and  ISO/SWS measurements of the radiance around 3.3\,$\mu$m have inferred significantly different methane concentrations. Here we aim to accurately model the \ch4 radiance at 3.3\,$\mu$m measured by ISO/SWS by using a comprehensive non-local thermodynamic equilibrium model and the most recent collisional rates measured in the laboratory for \ch4 to shed  new light onto the methane concentration in the upper atmosphere of Jupiter. These emission bands have been shown to present a peak contribution precisely at the $\mu$-bar level, hence directly probing the region of interest. 
We find that a high \ch4 concentration is necessary to explain the data, in contrast with the most recent analyses, and that the observations favour the lower limit of the latest laboratory measurements of the \ch4 collisional relaxation rates. Our results provide precise constraints on the composition and dynamics of the lower atmosphere of Jupiter.
}

   \keywords{keywords}

   \maketitle
%

\section{Introduction}
\label{introduction}

The energy budget in the stratosphere--thermosphere boundary of the atmosphere of Jupiter ($\sim10^{-3}$\,mbar) is largely governed by radiative processes \citep{appleby1990ch4, Moses2005, miller2013cooling}. In these layers methane (\ch4) is the most abundant spectroscopically active hydrocarbon species absorbing solar radiation. The incoming flux induces rotations and vibrations in the two bending and two stretching modes of \ch4  \cite[see e.g.][]{Garcia-Comas2011}, giving rise to strong emissions in the \ch4 near-IR bands that can be observed in the daylight atmosphere of Jupiter. 

At the $\mu$-bar level the methane abundance is determined by photolysis, which initiates a
 rich hydrocarbon chemistry \cite[see Fig.\,9 in][]{Moses2005}. At lower pressures (higher altitudes), the competition between eddy and molecular diffusion results in a decrease in hydrocarbon coolants (including \ch4) in the lower thermosphere, giving rise to the observed temperature inversion \citep{Yelle1996, Moses2005}.
In this context, accurately constraining the abundance of methane at the homopause is key to understanding the complex dynamics, the thermal structure, and the composition of the upper atmosphere of Jupiter. Unfortunately, the \ch4 concentration has not been measured in situ, but it is commonly inferred from its nadir radiance at 3.3\,$\mu$m in the daylight, which has been shown to probe the $10^{-1}$--$10^{-4}$\,mbar pressure layers \citep{Kim2009, Kim2014, kim2020temporal}.
Previous studies along this line  analysed the Infrared Space Observatory \cite[ISO,][]{Kessler1996, Graauw1996} measurements of \ch4 emission in the atmosphere of Jupiter using its Short Wavelength Spectrometer (SWS) \citep{Encrenaz1999, Drossart1999, Kim2014, panka2018modeling, kim2020temporal}.

In particular, \citet{Drossart1999} were able to nicely reproduce the ISO nadir spectrum by adjusting two collisional parameters in their non-local thermodynamic equilibrium (non-LTE) model and an eddy diffusion coefficient ($k_{zz}$) at the homopause in the range of (6--8)$\times 10^{6}$\,cm$^{2}$s$^{-1}$.
These $k_{zz}$ are similar to those of model~A in \cite{Moses2005}, and hence point to a rather high \ch4 volume mixing ratio (VMR) \citep[see e.g. Fig.\,14a in][]{Moses2005}. \citet{Drossart1999} did not discuss the sensitivity to different pressure-temperature (p-T) profiles, and they also  assumed collisional rates for the thermalisation of the solar pumped vibrational states by collisions with H$_2$ and for the inter-group redistribution of the absorbed energy which are about 20 and 10 times smaller, respectively, than those measured for the lower-energy states by \citet{Menard2005} a few years later.

Subsequently, \citet{Kim2014} analysed  the ISO/SWS Jupiter spectra finding \ch4 VMRs significantly lower than those derived by \cite{Drossart1999} and \citet{Moses2005}, 
with an upper limit of $\sim$6$\times 10^{-5}$ at 10$^{-3}$\,mbar. Although they did not report the populations of the emitting states, it is likely that this finding was caused by the rather low collisional rates they used (in comparison to those measured by \citealt{Menard2005}, see our Fig.\,\ref{rates}), which resulted in a slow thermalisation for the highly excited solar-pumped emitting states.

Interestingly, the nominal calculations of \citet{Kim2014} (models A and D) significantly underestimate the ISO radiance originating from the hot band ($v_{3}$+$v_{4}$$\rightarrow$$v_{4}$) centred  near 3.325\,$\mu$m. They did not focus specifically on this band, although its emission contains significant information that could help us constrain the collisional rates of  high-energy states in order  to better estimate the possible excitation of the $v_{3}$ state from the deactivation of the $v_{3}$+$v_{4}$ level, and possibly to retrieve the \ch4 concentration at pressure levels of $\sim$10\up{-2}\,mbar.

More recently, \citet{kim2020temporal} revised their analysis using the collisional rates from \citet{Menard2005}. The authors obtained significantly higher \ch4 abundances at the $\mu$-bar level than in their previous study, but still lower than those of \citet{Drossart1999}. However, we argue in Sect.\,\ref{discussion_rates} that this might be due to a misuse of the laboratory measurements of \citet{Menard2005}.

Here we re-examine the ISO/SWS observations of the \ch4 fluorescence near 3.3\,$\mu$m using a comprehensive non-LTE model for this molecule, also including the most recent measurements of the collisional rates by \cite{Menard2005} so as to fully exploit the information content of this spectrum. In particular, we our aim is to  distinguish between the divergent \ch4 VMR profiles derived so far from the nadir ISO measurements. We describe the ISO/SWS measurements and their calibration in Sect.\,\ref{observations}. In Sects.\,\ref{nlte_model} and \ref{modelling}  respectively we describe the non-LTE model and the model Jupiter atmospheres we used to obtain the results presented in Sect.\,\ref{results} and discussed in Sect.\,\ref{discussion}. Finally, in Sect.\,\ref{conclusions} we summarise the main conclusions of this work.

\section{Observations} 
\label{observations}
The nadir observations of the atmosphere of Jupiter were obtained on May 25, 1997 (UT), using the ISO/SWS spectrometer, which covers the 2.38\,--\,45.2\,$\mu$m spectral region \citep{Graauw1996}. We restricted the analysis to the 3.20\,--\,3.45\,$\mu$m interval (band 1D, grating order SW3), where \ch4 emits, using its grating mode with a resolving power of 1750--2150 \citep{Encrenaz1999}. The observations were taken by pointing the instrumental 14\,$\times$\,20\,arcsec$^2$ aperture at Jupiter's centre \citep{Drossart1999,Encrenaz1999}, while the solar zenith angle in the planet atmosphere was nearly 0\,deg. The aperture  was aligned  with its long-axis,  roughly parallel to the planet's polar axis. It covered latitudes in the range $\pm$30\textdegree\ and longitudes in the range $\pm$20\textdegree\ from the central meridian. The measured infrared spectrum is shown in
Fig.\,\ref{spectra1} and details about the reduction, calibration, and uncertainties are provided in \cite{Graauw1996} and \cite{Encrenaz1999}. In particular, \cite{Encrenaz1999} reported a 3$\sigma$ value of $\pm$1 jansky. In the following, we use the corresponding 1$\sigma$ uncertainties in units of \runits. The absolute flux accuracy is of the order of 20\%. In addition, we found that the best fit was obtained when applying a spectral offset of $-3\times$10$^{-5}$\,$\mu$m.

\section{The \ch4 non-LTE model}
\label{nlte_model}

The \ch4 non-LTE model used here is essentially that described in detail by \cite{Garcia-Comas2011}; it  is based on the Generic RAdiative traNsfer AnD non-LTE population Algorithm (GRANADA), originally built specifically for the Earth's atmosphere \citep{Funke2012}. This non-LTE model was successfully applied to explain the methane 3.3\,$\mu$m atmospheric emission observed by the Visible and Infrared Mapping Spectrometer instrument on board the Cassini spacecraft in the atmosphere of Titan. The bulk composition of Titan's atmosphere is of N$_2$, wheras the atmosphere of Jupiter contains mainly H$_2$ and He, and hence we  adapted our model to reflect the physical conditions of this gas giant and the different collisional partners and rates of the \ch4 vibrational states.

\begin{table*}[htbp]
\centering
\caption{\label{col_proc} Collisional processes included in the \ch4 non-LTE model.} 
\begin{tabular}{lllll}
\smallskipt
\hline\hline
\smallskipt
No. & Process  & Col. partner & Rate$^{\dag}$ (\cms) & Reference \\
\smallskipt
\hline\hline\smallskipt
1a &  CH$_4$($v_{1}$,$v_{3}$,$v_{b}$)+M\df  & M=H$_2$, He& $k_{g,a}$\,=\,$ 2.45\cdot 10^{-11}(0.237+2.58\cdot 10^{-3}T$) & 1 \cr 
  &  ~~~~CH$_4$($v_{1}$+1,$v_{3}$$-$1,$v_{b}$)+M & M=\ch4 & $k_{g,a,CH_4}$\,=\,$1.0\cdot 10^{-11}(1.157+7.48\cdot 10^{-3}T$)  & 1 \cr 
 1b &  CH$_4$($v_{s}$,$v_{2}$,$v_{4}$)+M \df  & M=H$_2$, He & $k_{g,b}$\,=\,$ 3.68\cdot 10^{-11} (0.313+2.32\cdot 10^{-3}T$) & 1 \cr 
  &  ~~~~CH$_4$($v_{s}$,$v_{2}$$-$1,$v_{4}$+1)+M & M=\ch4 & $k_{g,b,CH_4}$\,=\,$1.0\cdot 10^{-11}(1.146+18.9\cdot 10^{-3}T$)& 1 \cr 
1c &  CH$_4$($v_{3}$,$v_{b}$)+M \df & M=H$_2$, He & $k_{g,c}$\,=\,$ 3.53\cdot10^{-11}(-0.57+5.31\cdot10^{-3}T$) & 1 \cr 
  &  ~~~~CH$_4$($v_{3}$$-$1,$v_{b}$+$v_{2}$+$v_{4}$)+M  & M=\ch4 & $k_{g,c,CH_4}$\,=\,$1.0\cdot 10^{-11}(1.726+9.71\cdot 10^{-3}T$) & 1 \cr 
1d$^{\ast}$ &  CH$_4$($v_{3}$,$v_{b}$)+M \df & M=H$_2$, He, \ch4 &  $k_{g,d}$\,=\, $1.0\cdot10^{-12}(0.976+5.49\cdot10^{-3}T$) &  2 \cr
  &  ~~~~CH$_4$($v_{3}$$-$1,$v_{b}$+2$v_{4}$)+M  &   &   &   \cr
\smallskipt\hline\smallskipt
2 &  CH$^i_4$($v_{1}$,$v_{3}$,$v_{b}$)+M\df & M=H$_2$ &   $k_{vt,H_2}$\,=\,$v_{b}$\, 2.48$\cdot 10^{-14}\,\exp(9.11\cdot 10^{-3}T$) & 1 \cr 
  &  ~~~~CH$^i_4$($v_{1}$,$v_{3}$,$v_{b}$$-$1)+M & M=He & $k_{vt,He}$\,=\,$v_{b}$\, $6.37\cdot 10^{-16} \exp(0.01111\,T$) & 1 \cr 
 &    & M=\ch4 & $k_{vt,CH_4}$\,=\, $v_{b}$\, $10^{-16} \exp(0.0347\,T-5.14\cdot 10^{-5}T^2$) & 1 \cr 
\smallskipt\hline\smallskipt
3a &  \multicolumn{2}{l}{CH$_4$($v_{s}$,$v_{4}$)+\ch4 \df\ CH$_4$($v_{s}$,$v_{4}$$-$1)+\ch4($v_{4}$)} & $k_{vv,4}$\,= $v_{4}$ $1.1\cdot 10^{-11}$  & 1 \cr 
3b & \multicolumn{2}{l}{CH$_4$($v_{s}$,$v_{2}$)+\ch4 \df\ CH$_4$($v_{s}$,$v_{2}$$-$1)+\ch4($v_{2}$)} & $k_{vv,2}$\,=\,$v_{2}$ $1.1\cdot 10^{-11}$ & 3 \cr  
3c &  \multicolumn{2}{l}{CH$_4$($v_{3}$,$v_{b}$)+\ch4 \df\ CH$_4$($v_{3}$$-$1,$v_{b}$)+\ch4($v_{3}$)} & $k_{vv,3}$\,=\,$v_{3}$ $ 2.6\cdot10^{-12}$ & 1 \cr 
\smallskipt\hline\smallskipt
4a &  \multicolumn{2}{l}{CH$^{i}_4$($v_{s}$,$v_{b}$)+ CH$^{j}_4$ \df\ CH$^{i}_4$($v_{s}$,$v_{b}$$-$1)+ CH$^{j}_4$($v_{b}$=1)} & $k_{vv4,iso}$\,=\,$1.1\cdot10^{-11}$ & 3 \cr
4b &  \multicolumn{2}{l}{CH$^{i}_4$($v_{s}$,$v_{b}$)+ CH$^{j}_4$ \df\ CH$^{i}_4$($v_{s}$$-$1,$v_{b}$)+ CH$^{j}_4$($v_{s}$=1)} & $k_{vv3,iso}$\,=\,2.6\,$\cdot10^{-12}$ & 3 \cr
\smallskipt\hline\hline\smallskipt
\end{tabular}
\tablefoot{{$^{\dag}$Rate coefficient for the forward direction of the process. Backward values are computed assuming detailed balance \citep[see e.g. Eq.\,3.54 of][]{Lopez-Puertas2001}.} $v_{b}$ is the bending quantum ($v_{2}$ or $v_{4}$). $v_{s}$ is the stretching quantum ($v_{1}$ or $v_{3}$). 
$^{\ast}$Not considered in the nominal model. $i$ and $j$ refer to isotopologues $^{12}$\ch4 and $^{13}$\ch4, respectively. 
Rates at temperatures below 160\,K and above 350\,K are fixed to the values at those temperatures, respectively.
}
\tablebib{(1)~\citet{Menard2005}; (2) \citet{Hess1976}; (3) This work. 
}
\end{table*}

The energy levels and radiative transitions considered are illustrated in Fig.\,1 \citep[see also Fig.\,2 in][]{Garcia-Comas2011} and listed in Tables\,\ref{levels} and \ref{bands}. 
We considered four groups of energy levels (polyads) that involve the four fundamental modes of vibration of the main isotope of \ch4. Namely, the bending modes $v_{2}$ (symmetric) and $v_{4}$ (asymmetric), and the stretching modes $v_{1}$ (symmetric) and $v_{3}$ (assymetric). The grouping facilitated the computation of collisional processes involving many levels, but individual populations were computed separately for each  of them. That is, we did not assume the levels in a given group to be in LTE among themselves.  
In addition to the main $^{12}$\ch4 isotopologue levels, we also included the $v_{4}$, 2$v_{4}$, and $v_{3}$ states of the second most abundant $^{13}$\ch4 isotopologue.
The energy levels listed above are coupled by the collisional processes listed in Table \ref{col_proc} (discussed below) and, in addition, they are considered to be inter-connected by the vibrational bands listed in Table\,\ref{bands}. As some of these bands are rather weak, it was not necessary to take into account the exchange of photons between atmospheric layers due to radiative excitation or deexcitation. The bands for which radiative transfer was included are flagged  `RT' in Table\,\ref{bands}.

\begin{figure}
\includegraphics[angle=0, width=1.0\columnwidth]{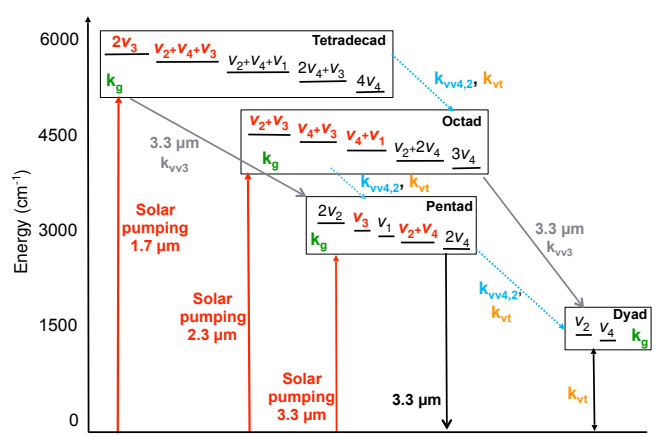}
\caption{Energy levels and groups, radiative transitions, and collisional processes considered in this work for the main isotope of \ch4. The solar-pumped energy levels are indicated in red. Adapted from \citet{Garcia-Comas2011}.} 
\label{model_sketch}
\end{figure}

The resulting system of equations is then composed of the statistical equilibrium equation for the energy levels involved and the radiative transfer equations for the corresponding vibrational bands. It was solved by a combination of the Curtis matrix and lambda iteration methods. For more details, see  Sect.\,9 in \cite{Funke2012}. The spectroscopic data used for  the calculation of the non-LTE populations described in this section and for the nadir radiance (see Sect.\,\ref{rad}) were mainly taken from the HITRAN 2016 compilation \citep{Gordon2017}. Some \ch4 bands in that compilation are incomplete as some predicted ro-vibrational lines included in previous HITRAN editions were removed. We replaced these bands by the corresponding ones of the HITRAN 2012 compilation \citep{Rothman2013}. In addition, some of the high-energy levels are connected through vibrational bands that are missing in those compilations and were taken from the MeCaSDa database \citep{Ba2013}. Table\,\ref{bands} lists all the bands with their Einstein coefficients and respective sources.

The excitation of the \ch4 levels in the stratosphere and thermosphere of Jupiter is triggered by solar pumping in the near-infrared. Table\,\ref{bands} indicates the bands where absorption of solar radiation was included. The spectral solar flux at  1\,AU included in this work for this spectral region was described in detail in \cite{Jurado-Navarro2015} and was adapted to the Sun--Jupiter distance at the time of the observations.

The collisional scheme implemented in this work (the nominal model) follows the laboratory measurements and subsequent analysis of \citet{Menard2005}, who studied collisions between the excited states of \ch4 with H$_2$, He, and  other \ch4 molecules.
They considered three main types of energy transfer processes: (i) intermode energy transfer; (ii) V--T (vibrational-thermal) energy transfer, and (iii) near-resonant vibration-vibration energy transfer processes. The intermode energy transfer processes redistribute the absorbed solar energy among the levels within a polyad through collisions with H$_2$, He, and \ch4. These processes comprise the energy transfer among the  (i) stretching modes ($v_{3}$\df$v_{1}$); (ii) bending modes ($v_{2}$\df$v_{4}$); and (iii) stretching and bending modes (2$v_{b}$\df$v_{s}$; see processes 1a, 1b, and 1 c, respectively, in Table\,\ref{col_proc}). The processes in (iii) can be of different types depending on the initial stretching mode, $v_{1}$ or $v_{3}$, and the final bending mode, 2$v_{4}$ or $v_{2}$+$v_{4}$. Here we follow \citet{Menard2005} and, in our nominal model, we consider only the $v_{3}$\flecha $v_{2}$+$v_{4}$ path (process 1c). Other authors \citep[e.g.][]{Hess1980, Kim2014} have considered that the main relaxation of the $v_{3}$ state occurs through $v_{3}$\flecha 2$v_{4}$ (process 1d). As the resulting populations of the \ch4 excited states strongly depend on the assumed relaxation path, we  studied both possibilities (see Sect.\,\ref{discussion}). 

The V-T energy transfer processes are responsible for the final relaxation of vibrational energy (pumped by absorption of solar radiation) into thermal energy. These processes are usually much slower than the intermode and V-V self-energy redistribution and involve the transitions where the transferred energy is the lowest possible (e.g. through the $v_{4}$ and $v_{2}$ bending quanta; see process~2 in Table\,\ref{col_proc}). We note  that these processes couple the energy levels among different polyads (Fig.\,\ref{model_sketch}). 

The near-resonant vibration-vibration energy transfer processes are assumed to occur only upon \ch4 self-collisions. These processes redistribute the energy among polyads by  exchanging one $v_{4}$ or one $v_{2}$ quanta, and by  exchanging one $v_{3}$ quantum (see processes 3a, 3b, and 3c in Table\,\ref{col_proc}). Similar vibration-vibration energy transfer processes between the two considered isotopologues of \ch4 are also included (processes 4a and 4b in Table\,\ref{col_proc}).
 
The values of the rates of the collisional processes are from \citet{Menard2005} and were measured at 193\,K and 296\,K. We fitted the expressions in Table\,\ref{col_proc} to the measured values. At temperatures below 160\,K and above 350\,K we considered the values at 160\,K and 350\,K, respectively. In the derivation of the measured rates, \citet{Menard2005} assumed that all $v_{3}$\flecha $v_{1}$ intermode transfer processes (i.e. irrespective of the $v_{b}$ excitation of the states) have the same rate coefficients for a given collisional partner.
Regarding the near resonant V--V processes, they are scaled by the factor predicted with the first-order perturbation theory for harmonic oscillators. The same approach was used for the rate coefficients of the V--T transfer processes (process 2 in Table\,\ref{col_proc}). 
In addition to the V--V transfer of $v_{4}$ in \citet{Menard2005}, we included a similar process with the exchange of a $v_{2}$ quanta for which we assumed the same value. Near-resonant V--V transfer processes for $v_{4}$ and $v_{3}$ exchanges between \ch4 isotopologues were also included in our scheme. For these we assumed the same rate coefficients measured by \citet{Menard2005} for similar processes in the main isotopologue. 
Processes 1--3 in Table\,\ref{col_proc} were also considered assuming the rate coefficients for both isotopologues.

\begin{figure}
\includegraphics[angle=90, width=1.0\columnwidth]{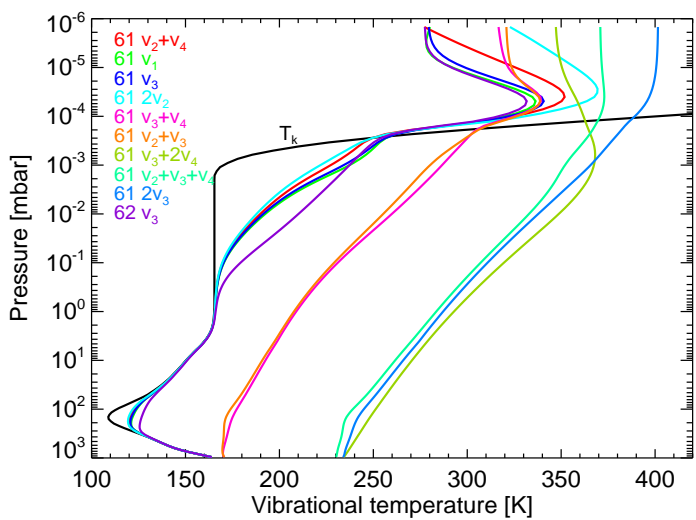}
\caption{Daytime (SZA=0\deg) vibrational temperatures of the main \ch4 energy levels contributing to the 3.3\,\um\ radiance as a function of pressure. The p-T$_1$ kinetic temperature profile, the VMR$_1$ abundance profile, and the nominal collisional rates of Table\,\ref{col_proc} were used. The major \ch4 isotopologue is labelled `61' and the second is labelled `62'.} 
\label{vt1}
\end{figure}

Figure~\ref{vt1} shows the resulting daytime (SZA=0$^{\circ}$) vibrational temperatures of the energy levels of \ch4 with energies close to and larger than 3.3\,$\mu$m for the nominal profile p-T$_1$, the \ch4 volume mixing ratio VMR$_1$ (see Sect.\,\ref{modelling}), and the nominal collisional rates of the non-LTE model (Table\,\ref{col_proc}). We observed that the $v_{3}$ level, the major contributor to the 3.3\,$\mu$m radiance, remains in LTE at pressures higher than $\sim$10$^{-1}$\,mbar, whereas the $v_{3}$+$v_{4}$ level is significantly far from LTE conditions in this region. At higher altitudes, $v_{3}$ departs from LTE and thus decouples from the kinetic temperature (T$_k$). Although both levels remain in non-LTE at lower pressures, the strong kinetic temperature inversion of p-T$_1$ at pressures below 10$^{-3}$\,mbar impacts their vibrational temperatures, which peak at $\sim$5$\times$10$^{-5}$\,mbar. At higher altitudes and at pressures lower than $\sim$10$^{-6}$\,mbar the T$_{vib}$ of both levels become isothermal since they are almost completely decoupled from T$_k$ due to the very low atmospheric density. 
The vibrational temperatures of the more energetic energy levels are significantly higher and more decoupled from LTE. This is mainly due to the high-frequency solar flux penetrating deeper in the atmosphere and due to the less efficient deexcitations for these high-energy levels. We note that the vibrational temperatures of these energy levels near 10$^{-3}$\,mbar depend very closely on the collisional rates (see Sect.\,\ref{k_tests}).

\section{Atmospheric modelling}
\label{modelling}

\subsection{Inputs for the \ch4 abundance and pressure-temperature structures}
\label{inputs_pt}

The measured radiance in a nadir sounding depends firstly and most importantly on the density of the emitting gas, \ch4 in our case, which dictates the atmospheric region where the emission emanates. Secondly, it depends on the population of the emitting levels. The latter are controlled by the kinetic temperature under LTE conditions, and by the vibrational temperatures under non-LTE, which might also depend, although more weakly, on the kinetic temperature profile itself. 

\begin{figure}
\includegraphics[angle=90, width=1.0\columnwidth]{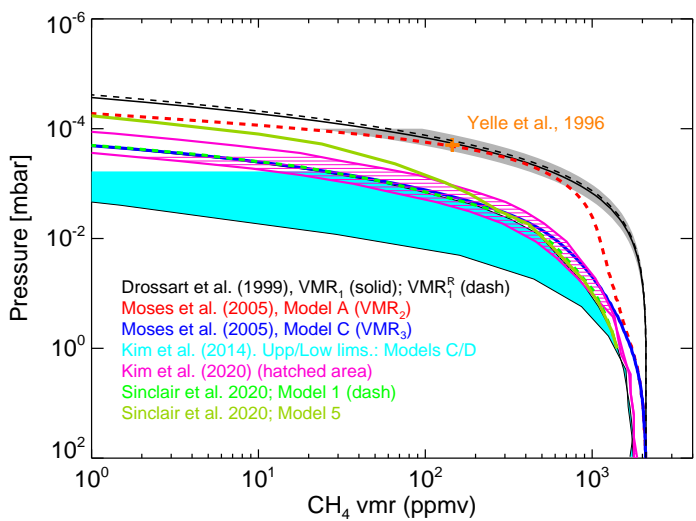}
\caption{\ch4 volume mixing ratio profiles considered in this work. The black dashed profile (VMR$_1^R$) is the \ch4 abundance that we derived in our analyses. The grey shaded area are the errors obtained by propagating the uncertainties of the collisional rates, temperature, and calibration error (see Sect.\,\ref{results}). The rest of the curves correspond to different profiles available in the literature, as labelled in the figure.}
\label{vmrs}
\end{figure}

We explored several abundance profiles of \ch4 available in the literature to explain the measured radiance. Figure\,\ref{vmrs} shows a compilation of the VMRs we used. VMR$_1$ coincides with the profile derived by \citet{Drossart1999} in the 10$^{-3}$ to 10$^{-4}$\,hPa region \citep[see Fig.\,14a of][]{Moses2005} and is built below assuming a vertically uniform eddy coefficient, with no (photo)chemical sources or sinks. 
A similar profile to VMR$_1$ (VMR$_1^R$, black dashed curve) was derived in this work (see Sect.\,\ref{results}) by propagating the uncertainties in the collisional rates of \citet{Menard2005}. More generally, the shaded area in grey corresponds to the range of VMRs we derived by considering additional sources of uncertainty, of which the absolute calibration error of the instrument is found to dominate (see Sect.\,\ref{sec_errors}).

\citet{Moses2005} proposed three different models, of which A and C present the highest and lowest \ch4 abundances, respectively. These profiles provided a reasonable fit to the C$_2$H$_6$ and C$_2$H$_2$ thermal infrared emissions observed by ISO \citep{fouchet2000jupiter} and differed by their assumed chemical reaction rates and pathways, and by their assumed eddy mixing coefficient (hence their homopause level). Model A featured an eddy mixing coefficient able to reproduce the \ch4 profile derived from UVS occultation \citep{Yelle1996} and the methane VMR inferred by \citet{Drossart1999} using the \ch4 fluorescent emissions observed by ISO, while their model C was optimised to reproduce the He airglow \citep{vervack1995jupiter}. For the chemistry, model C explored different reaction rates than model A so as to increase the net production rate of acetylene at high altitudes. In the following, we denote models A and C from \citeauthor{Moses2005} as VMR$_2$ and VMR$_3$, respectively.

{The shaded areas in cyan and pink in the figure show the range of VMR profiles derived by \cite{Kim2014} and \cite{kim2020temporal}, respectively, from the same ISO measurements. In the former work, the authors found rather low values for the abundance of methane, especially in their lower limit (model\,D). Even their upper limit, which is similar to the VMR$_3$ profile from \cite{Moses2005} and is also similar to the best-fitting profile in the updated analysis of \citet{kim2020temporal}}, is still significantly lower than the profile derived by \cite{Drossart1999} at pressures lower than $10$\,mbar, where the bulk emission at 3.3\,$\mu$m originates (see Fig.\,\ref{weights}). 

When comparing those profiles with previous measurements we find that the larger VMR described by \cite{Drossart1999} agrees very well with the stellar occultation measurements of \cite{Yelle1996}.
On the contrary, the profile obtained in \cite{Kim2014} and the updated abundance from \citet{kim2020temporal} fit rather well to the Lyman-$\alpha$ and He 584\,\AA\ airglow analysis reported by \cite{Moses2005}. The New Horizons ultraviolet stellar occultation measurements reported by \cite{Greathouse2010} also suggest a \ch4 VMR profile similar to model C of \cite{Kim2014}. However, the Galileo ASI analysis performed by \cite{Seiff1998} points to a high \ch4 in the 10$^{-3}-10^{-4}$\,mbar region.

\begin{figure}
\includegraphics[angle=90, width=1.0\columnwidth]{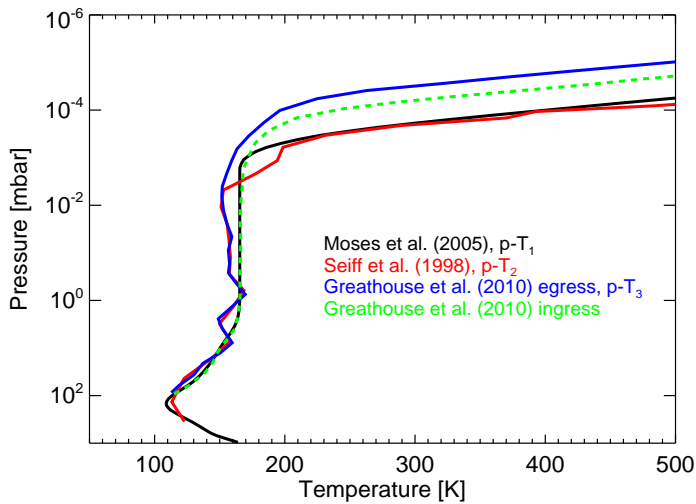}
\caption{Pressure--temperature profiles considered in this work. The colours correspond to different profiles available in the literature, as labelled in the figure.}
\label{profiles_pt}
\end{figure}

Under LTE conditions, the calculated nadir radiance depends directly on the kinetic temperature near the region where the emission emerges, roughly at $\sim$10$^{-2}-10^{-4}$\,mbar. In the case of the ISO measurements, however, the measured radiance seems to come from solar pumped excited \ch4 levels in non-LTE \citep[][see also below]{Drossart1999,Kim2014}. \cite{Drossart1999} did not discuss the effects of the kinetic temperature profile on the radiance, and \cite{Kim2009, Kim2014} found that the radiance was almost independent of the kinetic temperature. In contrast, in Sect.\,\ref{pt_tests} we show that the effects of T$_k$ on the radiance depend on the collisional rates used in the non-LTE model. Therefore, we considered in our analyses a range of p-T profiles available in the literature (see Fig.\,\ref{profiles_pt}). Specifically, our nominal profile (p-T$_1$) was computed by \citet{Moses2005}; p-T$_2$ was derived using the Galileo probe by \citet{Seiff1998}; and p-T$_3$ was obtained by \citet{Greathouse2010} by studying stellar occultation measurements of the New Horizons spacecraft taken during egress. As was discussed in \citet{Greathouse2010}, the density profile of the atmosphere was well constrained with the spacecraft, but the inference of the pressure-altitude information was very difficult due to uncertainties in the gravity. Thus, although the shape of the p-T$_3$ profile is correct, the absolute pressure-altitude reference is more uncertain.

These three profiles encompass rather well the expected range of temperatures near $\sim$10$^{-3}$\,mbar. 
However, they are rather different near $\sim$10$^{-3}$\,mbar, which might have a significant impact on the \ch4 nadir radiances. At lower pressures (higher altitudes), where the emerging nadir radiance is expected to be small, profiles p-T$_1$ and p-T$_2$ are very similar with both presenting high temperatures, whereas p-T$_3$ shows significantly lower values.

\subsection{Radiance calculation, contributing bands, and upward contributing radiances}
\label{rad}

We  simulated the nadir radiance calculations by using the Karlsruhe Optimised and Precise Radiative Transfer Algorithm \citep[KOPRA;][]{Stiller2002}, which is capable of computing radiances under non-LTE conditions. We used the \ch4 linelist presented in Sect.\,\ref{nlte_model} and the line shapes were modelled with a Voigt profile. The radiance was computed internally at a very high spectral resolution (0.0002\,cm$^{-1}$) and later convolved using a Gaussian kernel with a full width at half maximum of 2\,cm$^{-1}$, which corresponds to a resolving power of $\sim$1500 at 3.3\,$\mu$m.
We performed the calculations including the stronger bands near 3.3\,$\mu$m (see Table\,\ref{bands}), which comprises the following bands of the most abundant isotopologue $^{12}$\ch4: $v_{3}$\flecha ground, $v_{1}$\flecha ground, 2$v_{2}$\flecha ground, $v_{2}$+$v_{4}$\flecha ground, $v_{3}$+$v_{4}$\flecha $v_{4}$, $v_{2}$+$v_{3}$\flecha $v_{2}$, 2$v_{3}$\flecha $v_{3}$, $v_{3}$+2$v_{4}$\flecha 2$v_{4}$, $v_{2}$+$v_{3}$+$v_{4}$\flecha $v_{2}$+$v_{4}$, and the $v_{3}$\flecha ground band of the second most abundant isotopologue $^{13}$\ch4. In the figures we also used the HITRAN notation, with the codes `61' and `62' referring to $^{12}$\ch4 and $^{13}$\ch4, respectively.

\begin{figure}
\includegraphics[angle=90, width=1.0\columnwidth]{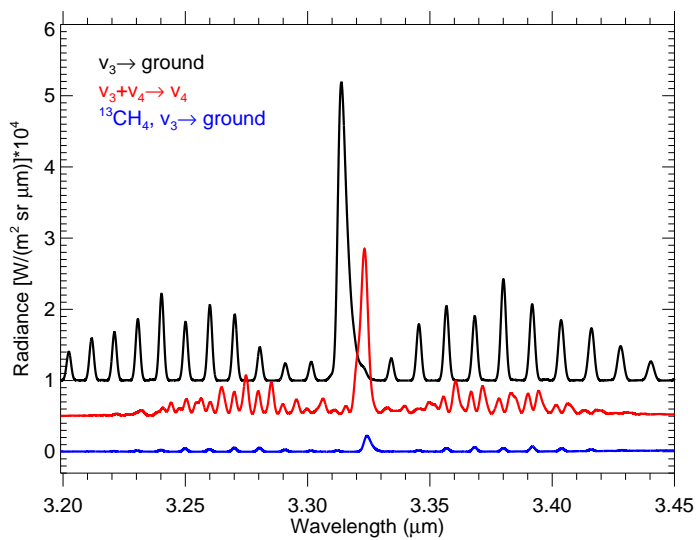}
\caption{
Contributions of \ch4 bands to the ISO nadir radiances near 3.3\,$\mu$m for the p-T$_1$ temperature, the \ch4 VMR$_1$ profile, and the nominal collisional rates (see Table\,\ref{col_proc}). A vertical offset of 0.5 units was included between the different radiances for clarity.} 
\label{contri_bands}
\end{figure}

In Figure \ref{contri_bands} we show the radiances of the major contributing bands to the nadir radiance for the p-T$_1$ and VMR$_1$ profiles and the nominal non-LTE model (Table\,\ref{col_proc}). 
We observed that the major contributions come from the fundamental $v_{3}$\flecha ground and the hot $v_{3}$+$v_{4}$\flecha$v_{4}$ bands of the major isotopologue. In addition, the Q branch of the fundamental $v_{3}$ band of the $^{13}$\ch4 isotopologue contributed slightly near 3.325\,$\mu$m.  The rest of the bands have a negligible contribution to the outgoing radiance. Although the contributions of the different bands strongly depend on the collisional relaxation rates, we observed that the fundamental and hot bands of the $^{12}$\ch4 isotopologue are always the major contributors for all the sets of collisional rates we tested (see Sect.\,\ref{results}).

\begin{figure}
\includegraphics[angle=90, width=1.0\columnwidth]{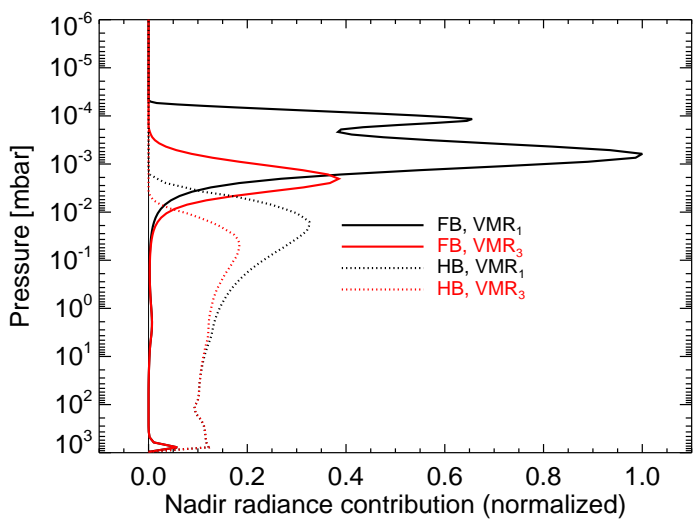}
\caption{Contributions to the upward radiance of the fundamental $v_{3}$ (FB, solid) and hot $v_{3}$+$v_{4}$ (HB, dashed) bands of \ch4 at the different pressure levels for the VMR$_1$ (black) and VMR$_3$ (red) profiles. The p-T$_1$ profile and the nominal non-LTE collisional rates of Table\,\ref{col_proc} were used in all cases. All values have been normalised to the maximum of the $v_{3}$ band for VMR$_1$.} 
\label{weights}
\end{figure}

We also computed the contribution to the nadir radiance of the different pressure levels for the stronger bands, the $v_{3}$\flecha ground and the hot $v_{3}$+$v_{4}$\flecha$v_{4}$ bands. The atmospheric region from which the radiance emanates depends significantly on the \ch4 concentration, so we computed the contribution functions for the significantly different \ch4 VMR$_1$ and VMR$_3$ profiles (see Fig.\,\ref{vmrs}).
The resulting contributions are shown in Fig.\,\ref{weights}. The radiance of the fundamental band mainly arises at pressure levels from 2$\times$10$^{-3}$ to 5$\times$10$^{-5}$\,mbar for the case of the larger VMR$_1$ profile. For the $v_{3}$+$v_{4}$ hot band, its emission originates at higher pressures (lower altitudes) from about 2$\times$10$^{-3}$\,mbar down to the lowest limit of our model. The significant contribution from high pressures is caused by the rather high excitation temperature (see Fig.\,\ref{vt_vmr}).
As expected, lower \ch4 concentrations (VMR$_3$) yield lower altitudes of emerging radiance, from 10$^{-2}$ to 2$\times$10$^{-4}$\,mbar for the fundamental band and from 10$^{-2}$ to the bottom of the model for the hot band. As the population of the emitting states of these bands increases very rapidly with altitude in these regions (see Fig.\,\ref{vt1}), the spectral radiance for the VMR$_3$ profile is then expected to be significantly smaller than for VMR$_1$.

\section{Results}
\label{results}

\begin{figure}
\includegraphics[angle=90, width=1.0\columnwidth]{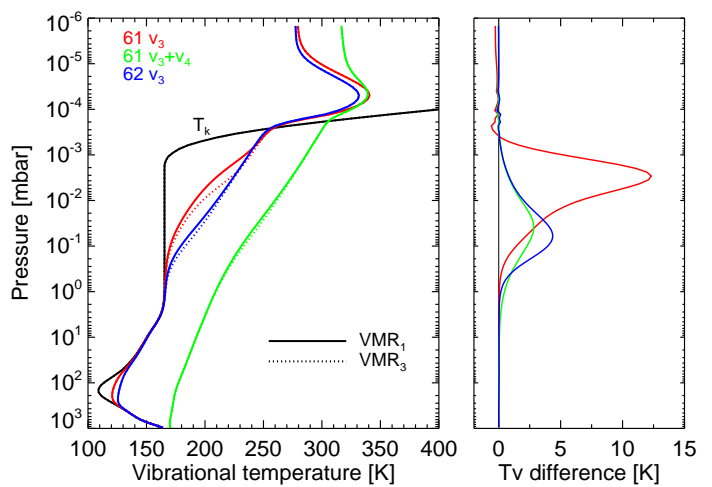}
\caption{Effects of the \ch4 VMR on the \ch4 vibrational temperatures for the main energy levels that contribute to the ISO 3.3\,\um\ radiance. The p-T$_1$ kinetic temperature profile and the nominal collisional rates were used.} 
\label{vt_vmr}
\end{figure}

We modelled the ISO/SWS nadir observations using the non-LTE model described above and the most recent collisional rates measured in the laboratory by \cite{Menard2005} aiming at constraining the \ch4 abundance. As we note in Sect.\,\ref{modelling}, \citet{Drossart1999} found a rather high \ch4 VMR profile, similar to model A from \citet{Moses2005} (see Fig.\,\ref{vmrs}), but more recent studies by \cite{Kim2014, kim2020temporal} derived a significantly smaller \ch4 abundance in the upper stratosphere of Jupiter (see cyan and pink shaded areas in Fig.\,\ref{vmrs}). However, from Fig.\,\ref{weights}, we would expect significantly different nadir radiances for the VMR$_1$ and VMR$_3$ profiles. Thus, we calculated the vibrational temperatures with the nominal non-LTE collisional rates of Table\,\ref{col_proc} for both VMR profiles using, in a first instance, p-T$_1$. The \tvib\ of the emitting energy levels for profile VMR$_1$ are shown in Fig.\,\ref{vt1}; and the effects of the different VMR profiles on the population of the major contributing levels are illustrated in Fig.\,\ref{vt_vmr}. We observed that the vibrational temperatures were higher for the lower VMR$_3$ profile at pressures in the range of 1--10$^{-3}$\,mbar. This is explained by the lower \ch4 column density, which allows the solar flux to penetrate deeper into the atmosphere, populating high-energy levels more efficiently than in the case of the VMR$_1$ profile.

\begin{figure}
\includegraphics[angle=90, width=1.0\columnwidth]{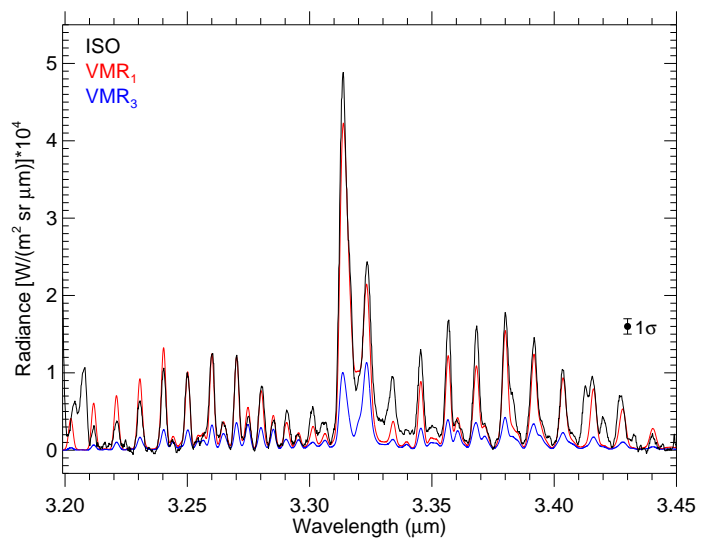}
\caption{Measured (black) and synthetic (red, VMR$_1$;  blue, VMR$_3$) spectra. The p-T$_1$ profile and the nominal collisional rates were used. The 1$\sigma$ noise error of the measurements is also shown.
} 
\label{spectra1}
\end{figure}

The radiances computed with those VMR profiles are shown in Fig.\,\ref{spectra1}. Here, we observed that using the VMR$_3$ profile largely underestimated the measured radiance. It is important to note that even if the \tvib\ values are higher for VMR$_3$, this does not compensate for the fact that the emitting region for this concentration occurs at considerably lower altitudes (see Fig.\,\ref{weights}) where the vibrational temperatures are much smaller.

Although the VMR$_1$ profile allowed us to obtain a reasonable fit, we still underestimated slightly the radiance from  the fundamental and the hot bands. Thus, we investigated the effects of the other two variables that mainly control the non-LTE populations of the emitting states and the nadir radiance, specifically the kinetic temperature profile and the collisional relaxation rates. In
addition, given the small radiances obtained using VMR$_3$, we  consider only the highest VMR$_1$ profile in the following.

\begin{figure}
\includegraphics[angle=90, width=1.0\columnwidth]{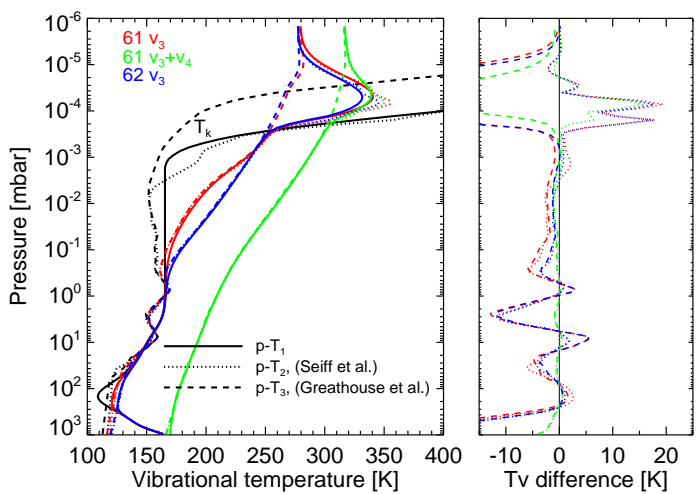}
\caption{Effects of the kinetic temperature profiles p-T$_1$, p-T$_2$, and p-T$_3$ on the \ch4 vibrational temperatures for the main energy levels that contribute to the ISO 3.3\,\um\ radiance. The VMR$_1$ profile and the nominal collisional rates were used.} 
\label{vt_seiff}
\end{figure}

\begin{figure}
\includegraphics[angle=90, width=1.0\columnwidth]{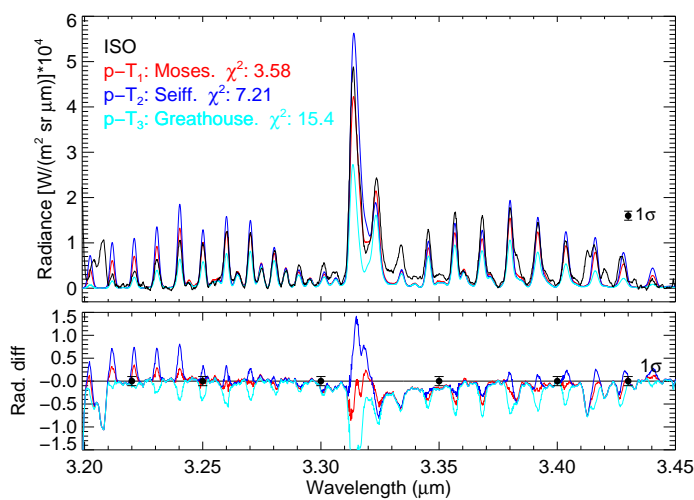}
\caption{Measured (black) and synthetic (red, p-T$_1$; blue, p-T$_2$;  cyan, p-T$_3$) spectra. The VMR$_1$ profile and the nominal collisional rates are   used. The 1$\sigma$ noise error of the measurements is also shown. As evidenced by the differences, p-T$_1$ provides the best fit to the observations.} 
\label{spectra_3pts}
\end{figure}

\subsection{Influence of temperature profiles} \label{pt_tests}

We  show above that the p-T$_1$ profile slightly underestimates the measured radiance near the Q branch for the nominal non-LTE model and the VMR$_1$ profile. Increasing the population of the $v_{3}$ level in the 10$^{-3}$--10$^{-4}$\,mbar region (see Figs.\,\ref{vt1} and \ref{weights}) requires a larger kinetic temperature at those pressure levels. For this  purpose, we computed the vibrational temperatures and model nadir radiances using the profile p-T$_2$ \citep{Seiff1998} and found a slight overestimation of the measured ISO spectrum in the Q branch of the fundamental band (see Figs.\,\ref{vt_seiff} and \ref{spectra_3pts}, respectively). This is mainly due to the increase in the \tvib\ of the $v_{3}$ level near 10$^{-4}$\,mbar  (see right panel in Fig.\,\ref{vt_seiff}). 
Furthermore, we found a significant overestimation of the radiance in the R branch of the fundamental $v_{3}$ band (see lower panel in Fig.\,\ref{spectra_3pts}). This suggests that the temperatures in p-T$_2$ at these pressure levels might be too large.
On the contrary, the $v_{3}$+$v_{4}$ emission is smaller when using p-T$_2$, which is caused by the lower kinetic temperature in the contribution region of this band (see Fig.\,\ref{weights}). In other words, although the population of $v_{3}$+$v_{4}$ is in non-LTE at all pressures, it is still affected by thermal relaxation, leading to a vibrational temperature decrease of a few kelvins (see right panel of Fig.\,\ref{vt_seiff}).

We also performed similar calculations for the colder p-T$_3$ profile still maintaining the same VMR$_1$ and the nominal collisional parameters. The results, as expected, yielded a significantly worse fit than for p-T$_1$ or p-T$_2$ due to the much lower vibrational temperatures of the $v_{3}$ and $v_{3}$+$v_{4}$ levels in their contributing layers (i.e. at pressures $\gtrsim$10$^{-4}$\,mbar and $\gtrsim$10$^{-3}$\,mbar, respectively).
 
Interestingly, \cite{Kim2014} found a weak dependence of their radiance calculations with the extreme kinetic temperatures profiles such as the low p-T$_3$ profile  \citep[egress,][]{Greathouse2010}, and the larger temperatures of p-T$_2$ \citep{Seiff1998}. As we discuss more in depth in Sect.\,\ref{k_tests}, this is likely due to their much lower collisional rates. 
We investigated the temperature dependence of the radiance for the nominal rates (see Fig.\,\ref{spectra_3pts}) and, in contrast with the results obtained by \citet{Kim2014}, we  observed a significant dependence of the vibrational temperatures on the kinetic temperature. This is a direct consequence of the higher rates of \citet{Menard2005} for the thermalisation of the solar pumped excited states that we used, which lead to a stronger coupling with the kinetic temperature.

\subsection{Influence of the collisional rates} \label{k_tests}

The other major source controlling the populations of the emitting levels, and hence the nadir radiance, is the collisional rates that deactivate the solar-pumped energy levels. The rate coefficients used by \cite{Drossart1999} and \cite{Kim2014} differ significantly from those measured in the laboratory by \citet{Menard2005}. \citet{kim2020temporal}  used the laboratory values. However, we argue in Sect.\,\ref{discussion_rates} that the authors misinterpreted the results of \citet{Menard2005} and used an incorrect rate for the collisional relaxation of the \ch4 excited states in collisions with H$_2$.

\begin{figure}
\includegraphics[angle=90, width=1.0\columnwidth]{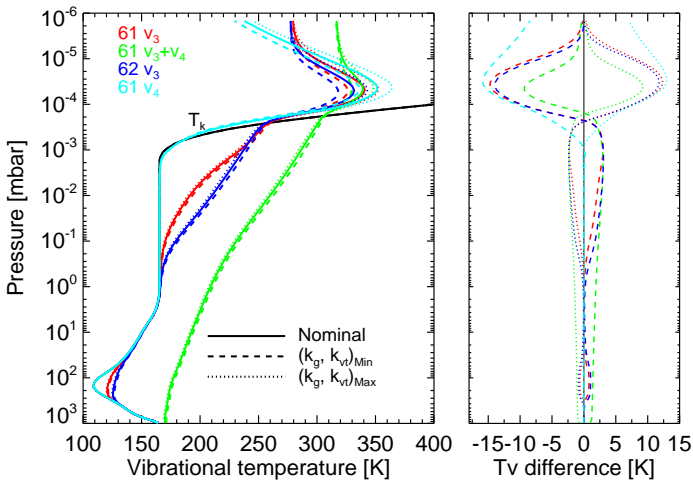}
\caption{Effects of perturbing the nominal \kg\ and \kvt\ rate coefficients by their errors ($\pm$25\%, factors of 0.75 and 1.25) on the \ch4 vibrational temperatures for the main energy levels that contribute to the ISO 3.3\,\um\ radiance. The VMR$_1^R$ and p-T$_1$ profiles were used.}
\label{vt_bestfit}
\end{figure}

\begin{figure}
\includegraphics[angle=90, width=1.0\columnwidth]{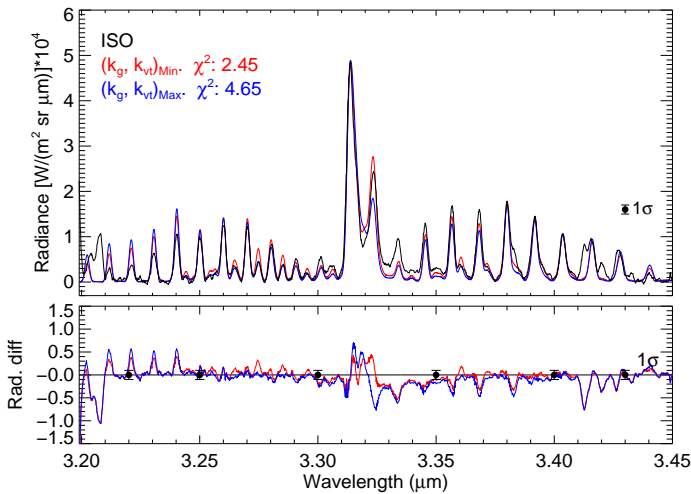}
\caption{{Effects of perturbing the nominal \kg\ and \kvt\ rate coefficients on the radiance by their errors. The measured spectrum is shown in black; the p-T$_1$ and VMR$_1^R$ profiles were used with the k$_{min}$ rates for the model in red, and with k$_{max}$ rates for the blue curve. The 1$\sigma$ noise error of the measurements is also shown. The relative differences are kept below $\sim$1\% in all cases.}} 
\label{spectra_bestfit}
\end{figure}

Since the rates of \citet{Menard2005} present an uncertainty of $\sim$20\%, here we have propagated them onto the derived \ch4 profile\footnote{In order to be more conservative, we   increased the uncertainty intervals to 25\%.}. To do so, we fixed the rates to their minimum (k$_{min}$) and maximum (k$_{max}$) values and derived the \ch4 profiles that provided the highest goodness of fit (lowest $\chi^2$) to the observations. Interestingly, we found that for both sets of rates the best fit is obtained when using the VMR$_1^R$ profile in Fig.\,\ref{vmrs}, which is very similar to that derived in \citet{Drossart1999}, here denoted VMR$_1$. Formally, the best-fitting abundance spans a very narrow area around the dashed curve, not much wider than the line thickness.
The reason for this narrow region is explained by the vibrational temperatures obtained for the k$_{min}$ and \kmax\ rates (see Fig.\,\ref{vt_bestfit}). 
For the lower \kmin\ rates, the solar pumped $v_{3}$ level is decoupled from the kinetic temperature, resulting in an increase of about 2--3\,K at p\,$\gtrsim$\,2$\times$10$^{-4}$\,mbar, and a decrease of up 10--15\,K in the upper layers; and the opposite occurs for the \kmax\ rates.
As illustrated in Fig.\,\ref{weights}, this band contributes to the nadir radiance mainly in the 10$^{-3}$--10$^{-4}$\,mbar region, where these increases and/or decreases of its vibrational temperature occur. Thus,  for \kmin\ the radiance of the region below $\sim10^{-4}$\,mbar   increases, while above that pressure it   decreases; and the opposite occurs for \kmax. Therefore, the overall result is that the total radiance hardly changes in the fundamental band (see Fig.\,\ref{spectra_bestfit}).
Remarkably, the \kmin\ case provided a better fit to ISO/SWS data  (we note the smaller $\chi^2$ values in Fig.\,\ref{spectra_bestfit}), including the three branches of the fundamental band and, particularly, the hot band; the latter is due to the larger population of its upper level for the \kmin\ rates below $\sim2\times10^{-4}$\,mbar  (Fig.\,\ref{vt_bestfit}). This could point to lower collisional rates than those reported by \citet{Menard2005}.

\begin{figure}
\includegraphics[angle=90, width=1.0\columnwidth]{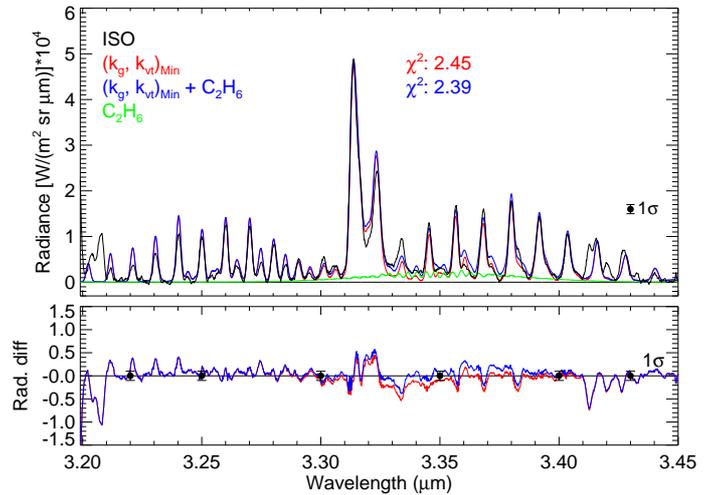}
\caption{
{Measured spectrum (black); model radiance using the p-T$_1$, the VMR$_1^R$ profile and the lower limit $k_{g,Min}$ and $k_{vt,Min}$ rates (red); estimated contribution of C$_2$H$_6$ (green); and sum of both contributions (blue). The 1$\sigma$ noise error of the measurements is also shown. The combined model fits slightly better the measurements (smaller $\chi^2$) as observed in the differences.}} 
\label{spectra_bestfit_c2h6}
\end{figure}

~\\
\subsection{Global error and overall fit}
\label{sec_errors}
We also estimated the errors in the \ch4 abundances propagated by the calibration uncertainty of 20\%. This source of error dominated over the uncertainties of the collisional rates discussed above and of the temperature (estimated to be about 5\%). 
In addition, we considered the errors in the line strengths of the fundamental and hot bands. For the former, the HITRAN 2016 compilation \citep{Gordon2017} reports an error of 5--10\% (uncertainty code\,=\,5). For the hot band there is no reported error in HITRAN, but the comparison of the total strengths in the HITRAN 2016 and MeCaSDa databases \citep{Ba2013} reveals just a 2\% difference. Hence, the spectroscopic errors for these measurements are much smaller than the other uncertainties. 
The total errors of including the uncertainties in these three parameters are shown by the shaded grey area in Fig.\,\ref{vmrs}.

The ISO radiances also present H$_3^+$ features at 3.414\,$\mu$m, which we did not include in our calculations. Additionally, the emission features at wavelengths shorter than 3.210\,$\mu$m were not considered in the goodness-of-fit evaluation since they present additional contributions from species other than \ch4 that were not studied here. 

Apart from these clear non-\ch4 emissions, there seems to be a systematic underestimation of the ISO radiance from 3.32 to 3.36\,\um, which we could not attribute to other \ch4 bands. We note that C$_2$H$_6$ has its \v7 band centred at this spectral region. We made a rough estimation of its possible contribution by using the C$_2$H$_6$ \v7 linelist of \citet{Villanueva2011}, assuming a C$_2$H$_6$ abundance profile ten times smaller than the \ch4 VMR$_1$ profile \citep{Nixon2007}, and adopting the \tvib\ of \ch4($v_{3}$) for the C$_2$H$_6$(\v7) state. The results show a better agreement with the ISO radiance (see Fig.\,\ref{spectra_bestfit_c2h6}), suggesting that C$_2$H$_6$ may also contribute significantly in this spectral region.

The fit to the ISO radiances still presents an overestimation of the lines of the R branch and an underestimation of the P branch lines. The reason behind these discrepancies is somewhat unclear. In principle, a different temperature profile might solve this issue since the emission from lines of different strengths emanates from different layers that might have different temperatures. However, this would not explain the differences obtained for lines with similar strengths in the two branches. We speculate that a possible explanation could be a drift in the calibration of the baseline of the instrument. In
addition, a background continuum emission from the lower atmosphere of Jupiter with a weak wavelength dependency, but modulated by the Planck function, could  partially explain the discrepancies.
Moreover, the P1 line is also systematically underestimated in all the tests performed. We investigated other \ch4 bands that could contribute in this region. The hot band $v_{2}$+$v_{3}$\flecha$v_{2}$ has a dense concentration of lines in the 3.32--3.34\,\um\ range that could enhance the emission of the P1 line. However, we found its contribution to be too weak  using the line distribution of HITRAN 2016 and the vibrational temperature computed with our model. It is thus uncertain at the moment which compound could enhance the P1 line emission to the levels observed with ISO.

\section{Discussion} \label{discussion}

\subsection{Collisional relaxation rates} 
\label{discussion_rates}
\begin{figure}
\includegraphics[angle=90, width=1.0\columnwidth]{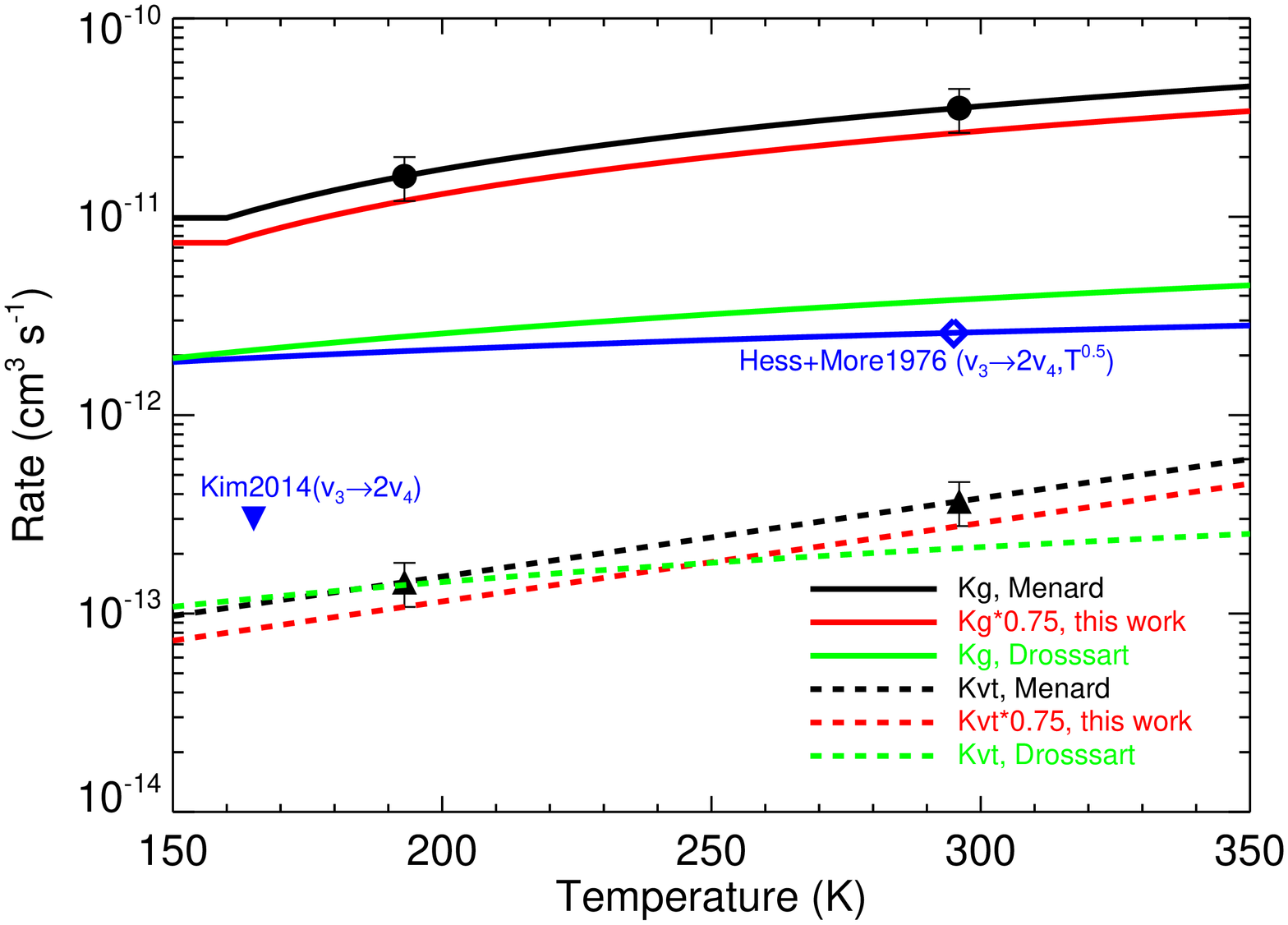}
\caption{Collisional rates for the deactivation of the $v_{3}$ state, $k_{g,c}$ or $k_{g,d}$ (solid curves), and for the relaxation of the $v_{4}$ state, \kvt\ (dashed curves). The rates measured in the laboratory by \cite{Menard2005} are shown with solid circles and triangles, in black, with their respective errors. The rates derived in \cite{Drossart1999} are shown with green curves. The \kvt\ rate measured by \cite{Hess1976} (blue open diamond) was used in \cite{Kim2014} with a $\sqrt{T}$-dependence and is indicated by the blue curve. The upper limit of the \kg\ rate used by \cite{Kim2014} to derive their lower limit model D for the \ch4 abundance is represented with a filled downward triangle in  blue. The rates derived in this work are shown in red.}\label{rates}
\end{figure}

The reduction of the collisional rates \kg\ and \kvt\ required to fit the ISO radiance with the VMR$_1^R$ profile is roughly consistent with the lower bound of \citet{Menard2005}. The rates derived or used in the previous analyses of \citet{Drossart1999} and \citet{Kim2014} were, however, significantly lower (see Fig.\,\ref{rates}), in particular that of the intermode energy transfer process, $k_{g,c}$, which is the major deactivation path for $v_{3}$.

\cite{Drossart1999}, who used a simpler \ch4 non-LTE model than ours but  {a very similar} abundance profile, found a rate for the relaxation of the symmetric-asymmetric $v_{1}$ and $v_{3}$ modes, $a_{vv}$($v_{3}$), of 95\,s$^{-1}\mu$bar$^{-1}$, which corresponds to 2.6\,$\times$10$^{-12}$\,\cms\ at a temperature of 200\,K. This rate coefficient is essentially the same measured by \cite{Hess1976} of 2.6$\times10^{-12}$\,\cms\ at 295\,K. On the contrary, it is about one order of magnitude smaller than that measured by \cite{Menard2005} and also than the value derived in this work (see Fig.\,\ref{rates}).

There are several reasons that could explain the different \kg\ collisional rates derived by \cite{Drossart1999} and in this work. On the one hand, the spectroscopic databases we used are richer and more complete than they were at the time of their publication. In
addition, and more importantly, the non-LTE schemes are significantly different, especially in their relaxation pathways. 

\begin{figure}
\includegraphics[angle=90, width=1.0\columnwidth]{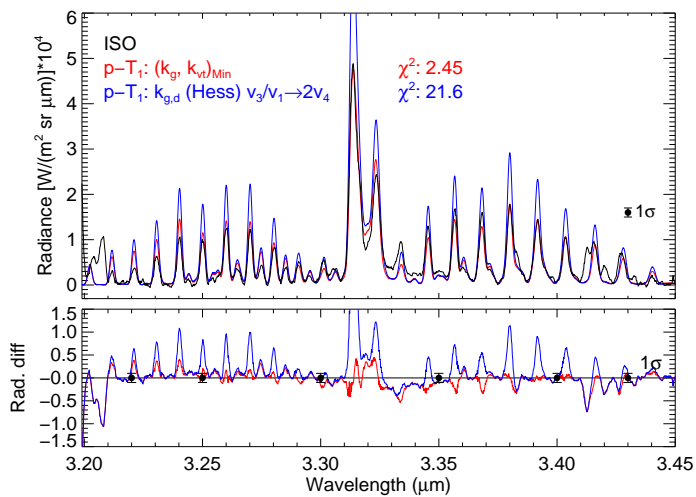}
\caption{Measured (black) and synthetic (red and blue) spectra. The red spectrum assumed the p-T$_1$ and VMR$_1^R$ profiles and the $k_{g,Min}$ and $k_{vt,Min}$ rates. The blue spectrum is obtained assuming the collisional rate $k_{g,d}$ instead of $k_{g,c}$ (see Table\,\ref{col_proc}) and also including  the relaxation of $v_{1}$ by the same process. The 1$\sigma$ noise error of the measurements is also shown.}
\label{spectra_bestfit_hess}
\end{figure}

\citet{Hess1976} found that the major relaxation path occurred through the $v_{3}$\flecha 2$v_{4}$ process, whereas \citet{Menard2005} (and this work) considered that the intermode energy exchange occurs primarily through $v_{3}$\flecha $v_{2}$+$v_{4}$. Although the energy of the 2$v_{4}$ level is smaller than that of $v_{2}$+$v_{4}$, this difference is probably not enough to justify the large discrepancy in the relaxation rates. We investigated this possibility by computing the radiances assuming that the $v_{3}$\flecha 2$v_{4}$ path (process $k_{g,d}$ in Table\,\ref{col_proc}) dominates and by including the relaxation of the $v_{1}$ mode following the same process. The results showed that the deactivation was still weak in this scenario, leading to large populations of the emitting levels and, consequently, to an overestimation of the radiance (see Fig.\,\ref{spectra_bestfit_hess}).

\cite{Kim2014} used the rate of \cite{Hess1976} to derive their upper limit for the \ch4 VMR. Furthermore, the final values they proposed were significantly lower, which resulted in even lower \ch4 abundances. In particular, their lower limit for the \ch4 VMR was obtained by using a very low relaxation rate for $v_{3}$ (`a rough mean of 3.5\,$\times$10$^{-16}$\,\cms\ and 3.0\,$\times$10$^{-13}$\,\cms'). The authors estimated this rate to overcome the lack of measurements at low temperatures at that time, and it was based on the relaxations of a different vibrational level, $v_{2}$, and of different partners, for example \ch4($v_{2}$) with H$_2$, and CD$_4$($v_{2}$) with H$_2$ measured at $\sim$165\,K. According to the measurement of \cite{Menard2005}, that value of $\sim$3.0\,$\times$10$^{-13}$\,\cms\ seems to be very small, and hence the low \ch4 abundance derived by \cite{Kim2014} is probably unlikely.

\cite{kim2020temporal} has more recently reanalysed the ISO measurements using the collisional rates measured by \cite{Menard2005}. They report the use of the laboratory rate coefficient $k_{v_3}$ from \citeauthor{Menard2005}, with values of 2.8$\pm$0.8\,$\times$10$^{-12}$\,\cms\ and 2.45$\pm$0.92\,$\times$10$^{-12}$\,\cms\ at 193\,K and 296\,K, respectively, for the vibrational relaxation of the \ch4 excited states in collisions with H$_2$. However, $k_{v_3}$ is actually the rate for the vibrational-vibrational exchange of $v_{3}$  quanta among \ch4\ levels in \ch4-\ch4 collisions \citep[see p. 3117, left column, in][]{Menard2005} and not the relaxation of the \ch4($v_{3}$,$v_{b}$) in collisions with H$_2$. Instead, we argue that the appropriate rate to use would have been the $k^{H_2}_{v_3\rightarrow v_2+v_4}$ rate in Table\,4 of \cite{Menard2005}, which is about a factor of ten higher than $k_{v_3}$. 

The effect of using a value for $k^{H_2}_{v_3\rightarrow v_2+v_4}$ (\kg) that is ten times smaller is shown in Fig.\,\ref{spectra_kgdiv10} for the VMR$_1$ profile. As expected, we obtained a large overestimation of the radiances. We also attempted to reproduce the \cite{kim2020temporal} results by using this low rate and the methane abundances that they derived (VMR$_3$ in Fig.\,\ref{vmrs}). The results are shown in Fig.\,\ref{spectra_kgdiv10_vmr3}. In this case we observed that the ISO radiances in the fundamental band are quite well reproduced, but the contribution of the hot band is largely overestimated. We note that the residuals very closely resemble the contribution of the hot band itself (see the red curve in Fig.\,\ref{contri_bands}). 
Therefore, their lower $k^{H_2}_{v_3\rightarrow v_2+v_4}$ rate is very likely the reason behind the significantly lower \ch4\ VMRs they obtained for the homopause, when compared to our results.

As for the vibrational-thermal (V--T) energy transfer (\kvt), which controls the thermalisation of the $v_{4}$ and $v_{2}$ bending quanta, we note that the values derived by \cite{Drossart1999} and those obtained here are very similar to those measured in the laboratory by \cite{Menard2005} and certainly within the measured 20\% uncertainties (dashed lines in Fig.\,\ref{rates}). Perhaps the temperature dependence of the rate coefficient at high temperatures in \citet{Drossart1999} is slightly weak.

\subsection{\ch4 abundance} 

We  show that the best fit to ISO/SWS measurements is obtained using the \ch4 abundance from the VMR$_1^R$ profile. This profile is very similar to the one used in \citet{Drossart1999} from 10$^{-3}$ to 10$^{-4}$\,mbar, and is also similar to other profiles reported in the literature, for example  to the profiles retrieved from the stellar occultation measurements by \citet{Yelle1996}, and to those obtained in the Galileo ASI analysis performed by \citet{Seiff1998}. On the contrary, it is significantly larger than the upper limit for the \ch4 VMR profile obtained by \citet{Kim2014} and the updated abundance reported in \cite{kim2020temporal}, both resulting from the analysis of the same ISO/SWS spectra. As we discuss in previous sections, a possible explanation for the discrepancy with the former is the use of very different collisional rates. As for the latter work, we argued in Sect.\,\ref{discussion_rates} that the authors might have misinterpreted the results of \citet{Menard2005} and used an incorrect rate for the relaxation of \ch4 levels in collisions with H$_2$, which lead them to a lower methane abundance than in this work.

Very recently, \cite{Sinclair2020} used 
ground-based IRTF-TEXES high-resolution spectra to compare the homopause location in auroral and non-auroral regions and inferred the \ch4 abundance near the homopause. They did not use the 3.3\,\um\ methane emission in the near-infrared, but studied mid-infrared \ch4 radiances arising from the $v_{4}$ level, and the CH$_3$ emission near 600\,\cm1. Figure\,\ref{vmrs} shows the range of the \ch4 VMR that these authors obtained. They found that at latitudes outside the auroral regions, their models 1 (low VMR) and 5 (medium VMR) yielded the best fits to their observed spectra at latitudes of 50\deg N and 68\deg N, respectively. The differences between the two models are relatively small, however. Their model 1 is very similar to the \ch4 VMR$_3$ profile of \cite{Moses2005}, and model 5 shows a higher abundance above $\sim10^{-3}$\,mbar, approaching the VMR$_1$ profile at lower pressures. Our best-fit \ch4 abundances in the VMR$_1^R$ profile are significantly higher than those of \cite{Sinclair2020} near the homopause, at pressures in the range of $10^{-2}$--$10^{-4}$\,mbar. A possible explanation for this discrepancy could be that the mid-infrared emission measurements they analysed might be probing slightly higher atmospheric heights than ISO/SWS data (see their Fig.\,4). In
addition, they considered the \ch4($v_{4}$) emission to be in LTE, whereas our model shows that the population of this band starts departing from LTE (with lower populations) at about 5$\times10^{-4}$\,mbar (see Fig.\ref{vt_bestfit}). Using the smallest non-LTE populations that we obtained for this level would result in higher retrieved \ch4 abundances, which would be in better agreement with the VMR$_1^R$ profile.

Accurately assessing the impact of our derived vertical profile of methane on the atmospheric energy budget, haze distribution, or kinetic temperature profile is challenging, as it would require a self-consistent modelling of the atmospheric radiative and chemical processes. However, our results do  suggest  that  methane  is  carried  to  higher layers than in some of  previous works (see Fig.\,3). This translates  into  a  stronger atmospheric  mixing  in  the  upper stratosphere and lower thermosphere.  In  addition, as the photolysis of methane and molecular diffusion deplete the high atmosphere of hydrocarbons coolants, the larger methane abundance hints at a slightly higher homopause and thermospheric base.

\section{Conclusions} \label{conclusions}

We have analysed ISO/SWS nadir measurements of \ch4 emission at 3.3\,$\mu$m arising from the homopause region in the atmosphere of Jupiter. By using a comprehensive non-LTE model to compute the population of the energy levels and the collisional relaxation rates (considering their uncertainties) reported by \citet{Menard2005}, we have presented the obtained abundance profile of \ch4, the kinetic and vibrational temperature structures, and the collisional relaxation parameters that allow us to best fit the observed infrared radiance.

Our analyses strongly favour a rather high volume mixing ratio of \ch4 at $\mu$-bar pressures. This is in good agreement with the results obtained by \citet{Drossart1999}, who analysed the same observations; with \citet{Yelle1996}, who studied stellar occultation using the Voyager Ultraviolet Spectrometer; and with  \cite{Seiff1998}, who performed the Galileo ASI analysis. On the contrary, our results do not agree with the low \ch4 abundance provided by \citet{Kim2014} from the ISO/SWS data, although the latter results are in line with those derived by \citet{Moses2005} from helium airglow spectra and the New Horizons ultraviolet stellar occultation measurements reported by \citet{Greathouse2010}. The very different collisional rate coefficients used in \citet{Kim2014}, lower than those of \citet{Menard2005}, are likely behind this discrepancy. Further, our derived methane abundance near the $\mu$-bar pressure level is significantly higher than the value reported in \citet{kim2020temporal}. However, an incorrect application of the collisional rates of \citet{Menard2005} in the latter work might be behind their lower \ch4 concentration.

We have propagated the uncertainties in the absolute calibration of ISO/SWS, in the collisional rates of \citet{Menard2005}, in the p-T profiles, and in the line strengths of the fundamental and hot bands onto the derived methane abundance. We find that the calibration uncertainties dominate over any other source of error. The final range of methane abundances we find still supports higher methane abundances than previously reported in the upper atmosphere of Jupiter. Interestingly, the minimum values of the collisional relaxation rates (i.e. a conservative reduction of 25\%) allowed us to best fit the observed radiance. In addition, we have found that the effects of the temperature profile on the radiance are dependent on the collisional rate coefficients. For the low rates used in \citet{Drossart1999} and \citet{Kim2014}, this dependence is negligible due to the strong decoupling. However, when using the rate coefficients of \cite{Menard2005}, the effect  becomes significant.

\begin{acknowledgements}
We thank the anonymous referee for their insightful reading of the manuscript and very useful suggestions. A.S.L. and I.S. acknowledge funding from the European Research Council (ERC) under the European Union's Horizon 2020 research and innovation program under grant agreement No 694513. IAA-CSIC authors acknowledge financial support from the Agencia Estatal de Investigaci\'on of the Ministerio de Ciencia, Innovaci\'on y Universidades through projects Ref. PID2019-110689RB-I00/AEI/10.13039/501100011033 and the Centre of Excellence ``Severo Ochoa'' award to the Instituto de Astrof\'isica de Andaluc\'ia (SEV-2017-0709).
\end{acknowledgements}

\bibliographystyle{aa} 
\bibliography{ref.bib}

\begin{appendix}

\section{ISO radiances using \kg\ rates 10 times smaller} \label{sec:rad_test_kg}

\begin{figure}[h]
\includegraphics[angle=90, width=1.0\columnwidth]{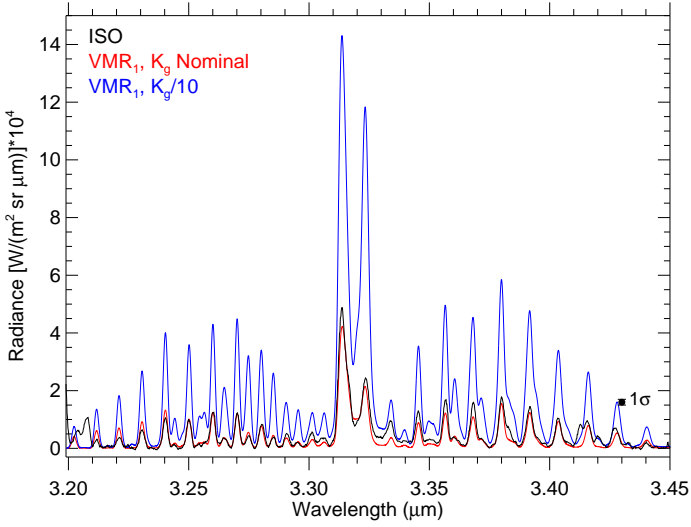}
\caption{Effects of using a \kg\ rate ten times lower than in Table\,\ref{col_proc}. Shown are the measured (black) and synthetic (red and blue) spectra. The red spectrum assumes the p-T$_1$ and VMR$_1$ profiles and the $k_{g}$ rate in Table\,\ref{col_proc}. The blue spectrum was obtained assuming a collisional rate $k_{g}$ that is  times smaller. The 1$\sigma$ noise error of the measurements is also shown.} 
\label{spectra_kgdiv10}
\end{figure}

\begin{figure}[ht]
\includegraphics[angle=90, width=1.0\columnwidth]{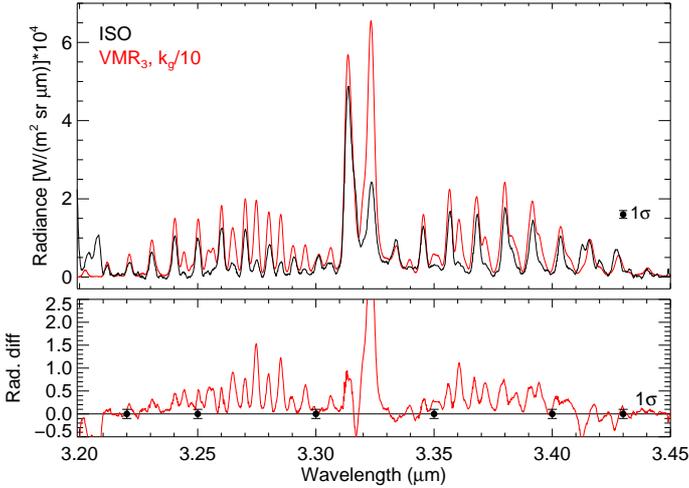}
\caption{Simulations using a \kg\ rate ten times smaller than in Table\,\ref{col_proc} and the \ch4 VMR$_3$ profile. The ISO measured (black) spectrum is also shown for reference. The residuals very closely resemble the hot band contribution (see red curve in Fig.\,\ref{contri_bands}). The 1$\sigma$ noise error of the measurements is also shown.} 
\label{spectra_kgdiv10_vmr3}
\end{figure}

\newpage

\section{\ch4 energy levels and vibrational bands} \label{sec:levels_bands}
Vibrational energy levels and bands considered in the non-LTE model.

%
\begin{table}[htbp]
\centering
\caption[]{\ch4 vibrational energy levels included in the model (updated from \citealt{Garcia-Comas2011}).} \label{levels} 
\tabcolsep 4 pt
\begin{tabular}{lrll}
\hline
Group &  ID$^{(a)}$ & Level & Energy (cm$^{-1}$) \\
\hline
 & 1 & Ground  &  0.00 \\
\hline
V$_{7.6}$  & 2 & $v_{4}$  &  1310.76 \\
                   & 3 & $v_{2}$  &  1533.33 \\
\hline
V$_{3.3}$  & 4 & 2$v_{4}$ &  2608.69 \\
                   & 6 & $v_{2}$+$v_{4}$ &  2838.26 \\
               & 7 & $v_{1}$ &  2916.48 \\
                   & 8 & $v_{3}$ & 3019.50 \\
                   & 9 & 2$v_{2}$ & 3064.48 \\
\hline
V$_{2.3}$  & 29 & 3$v_{4}$ & 3870.50 \\
                   & 44 & $v_{2}$+2$v_{4}$ & 4123.00 \\%
                   & 10 & $v_{1}$+$v_{4}$ & 4223.46 \\
                   & 11 & $v_{3}$+$v_{4}$ & 4321.67 \\
                   & 12 & $v_{2}$+$v_{3}$ & 4540.65 \\
\hline
V$_{1.7}$  & 56 & 4$v_{4}$ & 5161.10 \\%
                   & 30 & $v_{3}$+2$v_{4}$ & 5588.00 \\
                   & 102 & $v_{1}$+$v_{2}$+$v_{4}$ & 5775.00 \\%
                   & 103 & $v_{2}$+$v_{3}$+$v_{4}$ & 5861.00 \\%
                   & 13 & 2$v_{3}$ & 6024.28 \\
\hline
                   & 1 & Ground (62) &  0.00 \\
\hline
V$_{\rm 7.6, iso=2}$  & 2 & $v_{4}$  (62) &  1292.00 \\
V$_{\rm 3.3, iso=2}$  & 4 & 2$v_{4}$ (62) &  2588.16 \\
                              & 8 & $v_{3}$ (62) &  2999.06 \\
\hline
\end{tabular}
\tablefoot{
\tablefoottext{a}{HITRAN ID.}
}
\end{table}

\clearpage
\onecolumn

\begin{longtable}{lllrrrcl}
\caption{\label{bands} \ch4 vibrational bands included in the model.}\\
\hline\hline
Iso & Upper level & Lower level & (IDu) & (IDl) & $\bar\nu_0$ (cm$^{-1}$) & A$^{\dag}$ (s$^{-1}$) & Notes$^{\ddag}$ \\
\hline
\endfirsthead
\caption{continued.}\\
\hline\hline
Iso & Upper level & Lower level & (IDu) & (IDl) & $\bar\nu_0$ (cm$^{-1}$) & A$^{\dag}$  (s$^{-1}$) & Notes$^{\ddag}$ \\
\hline
\endhead
\hline
\endfoot
  1 &  2$v_{2}$         &  $v_{3}$          &     9 &     8 &      44.98 & 2.374e$-$06$^{a}$ &         -- \cr
  1 &  $v_{1}$          &  $v_{2}$+$v_{4}$      &     7 &     6 &      78.22 & 1.890e$-$05 &         -- \cr
  1 &  $v_{3}$          &  $v_{1}$          &     8 &     7 &     103.02 & 2.464e$-$05$^{a}$ &         -- \cr
  1 &  2$v_{2}$         &  $v_{1}$          &     9 &     7 &     148.00 & 6.095e$-$07$^{a}$ &         -- \cr
  1 &  $v_{3}$          &  $v_{2}$+$v_{4}$      &     8 &     6 &     181.24 & 2.686e$-$04$^{b}$ &         -- \cr
  1 &  $v_{2}$          &  $v_{4}$          &     3 &     2 &     222.57 & 5.539e$-$05 &         -- \cr
  1 &  2$v_{2}$         &  $v_{2}$+$v_{4}$      &     9 &     6 &     226.22 & 9.291e$-$05 &         -- \cr
  1 &  $v_{2}$+$v_{4}$      &  2$v_{4}$         &     6 &     4 &     229.57 & 6.766e$-$05 &         -- \cr
  1 &  $v_{1}$          &  2$v_{4}$         &     7 &     4 &     307.79 & 2.164e$-$04 &         -- \cr
  1 &  $v_{3}$          &  2$v_{4}$         &     8 &     4 &     410.81 & 8.929e$-$05 &         -- \cr
  1 &  2$v_{2}$         &  2$v_{4}$         &     9 &     4 &     455.79 & 2.006e$-$05 &         -- \cr
  1 &  4$v_{4}$         &  $v_{2}$+$v_{3}$      &    56 &    12 &     620.45 & 1.408e$-$05$^{a}$ &         -- \cr
  1 &  3$v_{4}$         &  2$v_{2}$         &    29 &     9 &     806.02 & 1.382e$-$05$^{a}$ &         -- \cr
  1 &  4$v_{4}$         &  $v_{3}$+$v_{4}$      &    56 &    11 &     839.43 & 3.527e$-$04$^{a}$ &         -- \cr
  1 &  3$v_{4}$         &  $v_{3}$          &    29 &     8 &     851.00 & 6.644e$-$05$^{a}$ &         -- \cr
  1 &  4$v_{4}$         &  $v_{1}$+$v_{4}$      &    56 &    10 &     937.64 & 1.016e$-$03$^{a}$ &         -- \cr
  1 &  3$v_{4}$         &  $v_{1}$          &    29 &     7 &     954.02 & 1.437e$-$04$^{a}$ &         -- \cr
  1 &  3$v_{4}$         &  $v_{2}$+$v_{4}$      &    29 &     6 &    1032.24 & 4.249e$-$03$^{a}$ &         -- \cr
  1 &  4$v_{4}$         &  $v_{2}$+2$v_{4}$     &    56 &    44 &    1038.10 & 2.393e$-$02$^{a}$ &         -- \cr
  1 &  $v_{3}$+2$v_{4}$     &  $v_{2}$+$v_{3}$      &    30 &    12 &    1047.35 & 4.783e$-$03$^{a}$ &         -- \cr
  1 &  $v_{2}$+2$v_{4}$     &  2$v_{2}$         &    44 &     9 &    1058.52 & 1.550e$-$03$^{a}$ &         -- \cr
  1 &  2$v_{4}$         &  $v_{2}$          &     4 &     3 &    1075.36 & 1.114e$-$03$^{a}$ &         -- \cr
  1 &  $v_{2}$+2$v_{4}$     &  $v_{3}$          &    44 &     8 &    1103.50 & 1.491e$-$03$^{a}$ &         -- \cr
  1 &  $v_{1}$+$v_{4}$      &  2$v_{2}$         &    10 &     9 &    1158.98 & 3.361e$-$03$^{a}$ &         -- \cr
  1 &  $v_{1}$+$v_{4}$      &  $v_{3}$          &    10 &     8 &    1203.96 & 5.998e$-$02$^{a}$ &         -- \cr
  1 &  $v_{2}$+2$v_{4}$     &  $v_{1}$          &    44 &     7 &    1206.52 & 8.738e$-$03$^{a}$ &         -- \cr
  1 &  $v_{1}$+$v_{2}$+$v_{4}$  &  $v_{2}$+$v_{3}$      &   102 &    12 &    1234.35 & 1.077e$-$01$^{a}$ &         -- \cr
  1 &  $v_{3}$+$v_{4}$      &  2$v_{2}$         &    11 &     9 &    1257.19 & 8.968e$-$02$^{a}$ &         -- \cr
  1 &  3$v_{4}$         &  2$v_{4}$         &    29 &     4 &    1261.81 & 6.621e$+$00$^{a}$ &         -- \cr
  1 &  $v_{3}$+2$v_{4}$     &  $v_{3}$+$v_{4}$      &    30 &    11 &    1266.33 & 3.826e$+$00$^{a}$ &         -- \cr
  1 &  $v_{2}$+2$v_{4}$     &  $v_{2}$+$v_{4}$      &    44 &     6 &    1284.74 & 4.675e$+$00$^{a}$ &         -- \cr
  1 &  4$v_{4}$         &  3$v_{4}$         &    56 &    29 &    1290.60 & 9.904e$+$00$^{a}$ &         -- \cr
  1 &  2$v_{4}$         &  $v_{4}$          &     4 &     2 &    1297.93 & 4.817e$+$00 &         RT \cr
  1 &  $v_{3}$+$v_{4}$      &  $v_{3}$          &    11 &     8 &    1302.17 & 2.236e$+$00$^{a}$ &         -- \cr
  1 &  $v_{2}$+$v_{4}$      &  $v_{2}$          &     6 &     3 &    1304.93 & 2.267e$+$00 &         -- \cr
  1 &  $v_{1}$+$v_{4}$      &  $v_{1}$          &    10 &     7 &    1306.98 & 2.229e$+$00$^{a}$ &         -- \cr
  1 &  $v_{4}$          &  Ground       &     2 &     1 &    1310.76 & 2.325e$+$00 &         RT \cr
  1 &  $v_{2}$+$v_{3}$+$v_{4}$  &  $v_{2}$+$v_{3}$      &   103 &    12 &    1320.35 & 2.391e$+$00$^{a}$ &         -- \cr
  1 &  $v_{3}$+2$v_{4}$     &  $v_{1}$+$v_{4}$      &    30 &    10 &    1364.54 & 1.522e$-$01$^{a}$ &         -- \cr
  1 &  $v_{1}$          &  $v_{2}$          &     7 &     3 &    1383.15 & 1.007e$-$01 &         RT \cr
  1 &  $v_{1}$+$v_{4}$      &  $v_{2}$+$v_{4}$      &    10 &     6 &    1385.20 & 4.126e$-$01$^{a}$ &         -- \cr
  1 &  $v_{3}$+$v_{4}$      &  $v_{1}$          &    11 &     7 &    1405.19 & 3.006e$-$02$^{a}$ &         -- \cr
  1 &  $v_{1}$+$v_{2}$+$v_{4}$  &  $v_{3}$+$v_{4}$      &   102 &    11 &    1453.33 & 1.746e$-$01$^{a}$ &         -- \cr
  1 &  $v_{3}$+2$v_{4}$     &  $v_{2}$+2$v_{4}$     &    30 &    44 &    1465.00 & 3.261e$-$01$^{a}$ &         -- \cr
  1 &  $v_{2}$+$v_{3}$      &  2$v_{2}$         &    12 &     9 &    1476.17 & 4.877e$-$01$^{a}$ &         -- \cr
  1 &  $v_{3}$+$v_{4}$      &  $v_{2}$+$v_{4}$      &    11 &     6 &    1483.41 & 3.318e$-$01$^{a}$ &         -- \cr
  1 &  2$v_{3}$         &  $v_{2}$+$v_{3}$      &    13 &    12 &    1483.63 & 6.624e$-$01$^{a}$ &         -- \cr
  1 &  $v_{3}$          &  $v_{2}$          &     8 &     3 &    1486.17 & 3.213e$-$01 &         -- \cr
  1 &  $v_{2}$+2$v_{4}$     &  2$v_{4}$         &    44 &     4 &    1514.31 & 9.028e$-$02$^{a}$ &         -- \cr
  1 &  $v_{2}$+$v_{3}$      &  $v_{3}$          &    12 &     8 &    1521.15 & 4.681e$-$02$^{a}$ &         -- \cr
  1 &  $v_{2}$+$v_{4}$      &  $v_{4}$          &     6 &     2 &    1527.50 & 5.846e$-$02 &         -- \cr
  1 &  2$v_{2}$         &  $v_{2}$          &     9 &     3 &    1531.15 & 1.034e$-$01 &         RT \cr
  1 &  $v_{2}$          &  Ground       &     3 &     1 &    1533.33 & 4.569e$-$02 &         RT \cr
  1 &  $v_{2}$+$v_{3}$+$v_{4}$  &  $v_{3}$+$v_{4}$      &   103 &    11 &    1539.33 & 1.400e$-$01$^{a}$ &         -- \cr
  1 &  $v_{1}$+$v_{2}$+$v_{4}$  &  $v_{1}$+$v_{4}$      &   102 &    10 &    1551.54 & 1.071e$-$01$^{a}$ &         -- \cr
  1 &  $v_{1}$          &  $v_{4}$          &     7 &     2 &    1605.72 & 3.004e$-$02 &         -- \cr
  1 &  $v_{1}$+$v_{4}$      &  2$v_{4}$         &    10 &     4 &    1614.77 & 1.029e$-$01$^{a}$ &         -- \cr
  1 &  $v_{2}$+$v_{3}$      &  $v_{1}$          &    12 &     7 &    1624.17 & 2.907e$-$02$^{a}$ &         -- \cr
  1 &  $v_{2}$+$v_{3}$+$v_{4}$  &  $v_{1}$+$v_{4}$      &   103 &    10 &    1637.54 & 9.270e$-$02$^{a}$ &         -- \cr
  1 &  $v_{1}$+$v_{2}$+$v_{4}$  &  $v_{2}$+2$v_{4}$     &   102 &    44 &    1652.00 & 1.247e$-$01$^{a}$ &         -- \cr
  1 &  $v_{2}$+$v_{3}$      &  $v_{2}$+$v_{4}$      &    12 &     6 &    1702.39 & 8.252e$-$01$^{a}$ &         -- \cr
  1 &  2$v_{3}$         &  $v_{3}$+$v_{4}$      &    13 &    11 &    1702.61 & 1.335e$+$00$^{a}$ &         -- \cr
  1 &  $v_{3}$          &  $v_{4}$          &     8 &     2 &    1708.74 & 8.407e$-$01 &         -- \cr
  1 &  $v_{3}$+$v_{4}$      &  2$v_{4}$         &    11 &     4 &    1712.98 & 1.151e$+$00$^{a}$ &         -- \cr
  1 &  $v_{3}$+2$v_{4}$     &  3$v_{4}$         &    30 &    29 &    1717.50 & 1.176e$+$00$^{a}$ &         -- \cr
  1 &  $v_{2}$+$v_{3}$+$v_{4}$  &  $v_{2}$+2$v_{4}$     &   103 &    44 &    1738.00 & 1.096e$+$00$^{a}$ &         -- \cr
  1 &  2$v_{2}$         &  $v_{4}$          &     9 &     2 &    1753.72 & 4.775e$-$02 &         -- \cr
  1 &  2$v_{3}$         &  $v_{1}$+$v_{4}$      &    13 &    10 &    1800.82 & 5.262e$-$02$^{a}$ &         -- \cr
  1 &  2$v_{3}$         &  $v_{2}$+2$v_{4}$     &    13 &    44 &    1901.28 & 1.915e$-$01$^{a}$ &         -- \cr
  1 &  $v_{1}$+$v_{2}$+$v_{4}$  &  3$v_{4}$         &   102 &    29 &    1904.50 & 4.939e$-$02$^{a}$ &         -- \cr
  1 &  $v_{2}$+$v_{3}$      &  2$v_{4}$         &    12 &     4 &    1931.96 & 6.019e$-$03$^{a}$ &         -- \cr
  1 &  $v_{2}$+$v_{3}$+$v_{4}$  &  3$v_{4}$         &   103 &    29 &    1990.50 & 1.708e$-$02$^{a}$ &         -- \cr
  1 &  4$v_{4}$         &  $v_{3}$          &    56 &     8 &    2141.60 & 2.600e$-$03$^{a}$ &         -- \cr
  1 &  2$v_{3}$         &  3$v_{4}$         &    13 &    29 &    2153.78 & 1.193e$-$01$^{a}$ &         -- \cr
  1 &  4$v_{4}$         &  $v_{1}$          &    56 &     7 &    2244.62 & 2.863e$-$05$^{a}$ &         -- \cr
  1 &  4$v_{4}$         &  $v_{2}$+$v_{4}$      &    56 &     6 &    2322.84 & 2.593e$-$03$^{a}$ &         -- \cr
  1 &  3$v_{4}$         &  $v_{2}$          &    29 &     3 &    2337.17 & 7.914e$-$04$^{a}$ &         -- \cr
  1 &  $v_{3}$+2$v_{4}$     &  2$v_{2}$         &    30 &     9 &    2523.52 & 1.024e$-$02$^{a}$ &         -- \cr
  1 &  4$v_{4}$         &  2$v_{4}$         &    56 &     4 &    2552.41 & 3.281e$-$01$^{a}$ &         -- \cr
  1 &  3$v_{4}$         &  $v_{4}$          &    29 &     2 &    2559.74 & 1.502e$-$01 &         -- \cr
  1 &  $v_{3}$+2$v_{4}$     &  $v_{3}$          &    30 &     8 &    2568.50 & 1.112e$-$01$^{a}$ &         -- \cr
  1 &  $v_{2}$+2$v_{4}$     &  $v_{2}$          &    44 &     3 &    2589.67 & 5.780e$-$02$^{a}$ &         -- \cr
  1 &  2$v_{4}$         &  Ground       &     4 &     1 &    2608.69 & 5.438e$-$02 &    RT, Sun \cr
  1 &  $v_{3}$+2$v_{4}$     &  $v_{1}$          &    30 &     7 &    2671.52 & 6.977e$-$03$^{a}$ &         -- \cr
  1 &  $v_{1}$+$v_{4}$      &  $v_{2}$          &    10 &     3 &    2690.13 & 4.369e$-$03 &         -- \cr
  1 &  $v_{1}$+$v_{2}$+$v_{4}$  &  2$v_{2}$         &   102 &     9 &    2710.52 & 2.226e$-$01$^{a}$ &         -- \cr
  1 &  $v_{3}$+2$v_{4}$     &  $v_{2}$+$v_{4}$      &    30 &     6 &    2749.74 & 2.137e$-$01$^{a}$ &         -- \cr
  1 &  $v_{1}$+$v_{2}$+$v_{4}$  &  $v_{3}$          &   102 &     8 &    2755.50 & 2.959e$-$01$^{a}$ &         -- \cr
  1 &  $v_{3}$+$v_{4}$      &  $v_{2}$          &    11 &     3 &    2788.34 & 2.706e$-$02 &         -- \cr
  1 &  $v_{2}$+$v_{3}$+$v_{4}$  &  2$v_{2}$         &   103 &     9 &    2796.52 & 9.335e$-$02$^{a}$ &         -- \cr
  1 &  $v_{2}$+2$v_{4}$     &  $v_{4}$          &    44 &     2 &    2812.24 & 5.756e$-$01 &         -- \cr
  1 &  $v_{2}$+$v_{4}$      &  Ground       &     6 &     1 &    2838.26 & 3.897e$-$01 &    RT, Sun \cr
  1 &  $v_{2}$+$v_{3}$+$v_{4}$  &  $v_{3}$          &   103 &     8 &    2841.50 & 8.059e$-$01$^{a}$ &         -- \cr
  1 &  $v_{1}$+$v_{2}$+$v_{4}$  &  $v_{1}$          &   102 &     7 &    2858.52 & 4.928e$-$01$^{a}$ &         -- \cr
  1 &  $v_{1}$+$v_{4}$      &  $v_{4}$          &    10 &     2 &    2912.70 & 5.254e$-$01 &         -- \cr
  1 &  $v_{1}$          &  Ground       &     7 &     1 &    2916.48 & 5.539e$-$03 &         RT \cr
  1 &  $v_{1}$+$v_{2}$+$v_{4}$  &  $v_{2}$+$v_{4}$      &   102 &     6 &    2936.74 & 1.321e$+$00$^{a}$ &         -- \cr
  1 &  $v_{2}$+$v_{3}$+$v_{4}$  &  $v_{1}$          &   103 &     7 &    2944.52 & 1.059e$+$00$^{a}$ &         -- \cr
  1 &  2$v_{3}$         &  2$v_{2}$         &    13 &     9 &    2959.80 & 4.295e$-$01$^{a}$ &         -- \cr
  1 &  $v_{3}$+2$v_{4}$     &  2$v_{4}$         &    30 &     4 &    2979.31 & 1.945e$+$01$^{a}$ &        Sun \cr
  1 &  2$v_{3}$         &  $v_{3}$          &    13 &     8 &    3004.78 & 4.554e$+$01$^{a}$ &        Sun \cr
  1 &  $v_{2}$+$v_{3}$      &  $v_{2}$          &    12 &     3 &    3007.32 & 2.251e$+$01 &         RT \cr
  1 &  $v_{3}$+$v_{4}$      &  $v_{4}$          &    11 &     2 &    3010.91 & 2.374e$+$01 &         RT \cr
  1 &  $v_{3}$          &  Ground       &     8 &     1 &    3019.50 & 2.493e$+$01 &    RT, Sun \cr
  1 &  $v_{2}$+$v_{3}$+$v_{4}$  &  $v_{2}$+$v_{4}$      &   103 &     6 &    3022.74 & 2.383e$+$01$^{a}$ &         -- \cr
  1 &  2$v_{2}$         &  Ground       &     9 &     1 &    3064.48 & 7.446e$-$02 &         RT \cr
  1 &  2$v_{3}$         &  $v_{1}$          &    13 &     7 &    3107.80 & 6.919e$-$01$^{a}$ &         -- \cr
  1 &  $v_{1}$+$v_{2}$+$v_{4}$  &  2$v_{4}$         &   102 &     4 &    3166.31 & 9.926e$-$01$^{a}$ &         -- \cr
  1 &  2$v_{3}$         &  $v_{2}$+$v_{4}$      &    13 &     6 &    3186.02 & 3.714e$+$00$^{a}$ &         -- \cr
  1 &  $v_{2}$+$v_{3}$      &  $v_{4}$          &    12 &     2 &    3229.89 & 9.060e$-$02 &         -- \cr
  1 &  $v_{2}$+$v_{3}$+$v_{4}$  &  2$v_{4}$         &   103 &     4 &    3252.31 & 2.348e$-$01$^{a}$ &         -- \cr
  1 &  2$v_{3}$         &  2$v_{4}$         &    13 &     4 &    3415.59 & 1.262e$+$00$^{a}$ &         -- \cr
  1 &  4$v_{4}$         &  $v_{2}$          &    56 &     3 &    3627.77 & 3.688e$-$04$^{a}$ &         -- \cr
  1 &  4$v_{4}$         &  $v_{4}$          &    56 &     2 &    3850.34 & 1.477e$-$01$^{a}$ &         -- \cr
  1 &  3$v_{4}$         &  Ground       &    29 &     1 &    3870.50 & 3.617e$-$02 &    RT, Sun \cr
  1 &  $v_{3}$+2$v_{4}$     &  $v_{2}$          &    30 &     3 &    4054.67 & 1.448e$-$02$^{a}$ &         -- \cr
  1 &  $v_{2}$+2$v_{4}$     &  Ground       &    44 &     1 &    4123.00 & 2.829e$-$02 &         -- \cr
  1 &  $v_{1}$+$v_{4}$      &  Ground       &    10 &     1 &    4223.46 & 1.567e$+$00 &    RT, Sun \cr
  1 &  $v_{1}$+$v_{2}$+$v_{4}$  &  $v_{2}$          &   102 &     3 &    4241.67 & 1.254e$+$00$^{a}$ &         -- \cr
  1 &  $v_{3}$+2$v_{4}$     &  $v_{4}$          &    30 &     2 &    4277.24 & 1.224e$+$00$^{a}$ &         -- \cr
  1 &  $v_{3}$+$v_{4}$      &  Ground       &    11 &     1 &    4321.67 & 5.222e$-$01 &    RT, Sun \cr
  1 &  $v_{2}$+$v_{3}$+$v_{4}$  &  $v_{2}$          &   103 &     3 &    4327.67 & 7.700e$-$01$^{a}$ &         -- \cr
  1 &  $v_{1}$+$v_{2}$+$v_{4}$  &  $v_{4}$          &   102 &     2 &    4464.24 & 1.298e$-$01$^{a}$ &         -- \cr
  1 &  2$v_{3}$         &  $v_{2}$          &    13 &     3 &    4490.95 & 2.979e$-$01$^{a}$ &         -- \cr
  1 &  $v_{2}$+$v_{3}$      &  Ground       &    12 &     1 &    4540.65 & 2.121e$-$01 &    RT, Sun \cr
  1 &  $v_{2}$+$v_{3}$+$v_{4}$  &  $v_{4}$          &   103 &     2 &    4550.24 & 2.013e$-$01$^{a}$ &         -- \cr
  1 &  2$v_{3}$         &  $v_{4}$          &    13 &     2 &    4713.52 & 1.094e$-$02 &         -- \cr
  1 &  4$v_{4}$         &  Ground       &    56 &     1 &    5161.10 & 1.007e$-$03 &         -- \cr
  1 &  $v_{3}$+2$v_{4}$     &  Ground       &    30 &     1 &    5588.00 & 2.854e$-$02$^{a}$ &  Sun \cr
  1 &  $v_{1}$+$v_{2}$+$v_{4}$  &  Ground       &   102 &     1 &    5775.00 & 3.419e$-$03$^{a}$ &         -- \cr
  1 &  $v_{2}$+$v_{3}$+$v_{4}$  &  Ground       &   103 &     1 &    5861.00 & 5.474e$-$02$^{a}$ &        Sun \cr
  1 &  2$v_{3}$         &  Ground       &    13 &     1 &    6024.28 & 3.289e$-$01$^{b}$ &  Sun \cr
  2 &  $v_{4}$          &  Ground       &     2 &     1 &    1292.00 & 2.055e$+$00 &         RT \cr
  2 &  2$v_{4}$         &  $v_{4}$          &     4 &     2 &    1296.16 & 4.632e$+$00 &         RT \cr
  2 &  2$v_{4}$         &  Ground       &     4 &     1 &    2588.16 & 5.133e$-$02 &    RT, Sun \cr
  2 &  $v_{3}$          &  Ground       &     8 &     1 &    2999.06 & 2.298e$+$01 &    RT, Sun \cr
\end{longtable}
\tablefoot{$^{\dag}$\,Einstein coefficient at 250~K calculated from the HITRAN 2016 edition \citep{Gordon2017}. $^{a}$\,Calculated from the MeCaSDa database \citep{Ba2013}. $^{b}$\,Calculated from the HITRAN 2012 database \citep{Rothman2013}. $^{\ddag}$\,RT: Full radiative transfer among atmospheric layers is included. Sun: The absorption of solar radiation in this band is important.}

\end{appendix}

\end{document}